\documentclass[runningheads]{llncs}
\usepackage[T1]{fontenc}
\usepackage{tabularx}
\usepackage{multirow}
\usepackage{amsmath}
\usepackage{mathptmx}
\usepackage{helvet}
\usepackage{courier}
\usepackage{type1cm}         
\usepackage{makeidx}         
\usepackage{graphicx}        
\usepackage{multicol}        
\usepackage[bottom]{footmisc}
\usepackage{newtxtext}       %
\usepackage[varvw]{newtxmath}       
\usepackage{hyperref}
\usepackage[sort&compress,numbers]{natbib}
\usepackage{subfig}
\usepackage{verbatim}
\usepackage{rotating}
\usepackage{xcolor}
\usepackage{tikz}
\usepackage{setspace}

\usepackage{musicography} 

\newcommand{\doublesharp}{\musDoubleSharp}
\newcommand{\doubleflat}{\musDoubleFlat}

\newcommand{\tc}{F}

\usetikzlibrary {arrows.meta,shapes.misc}
\date{}

\begin{document}
\author{David Meredith}
\title{Geometric Representations for Transformed Pattern Matching in Music}
\titlerunning{Geometric Representations for Music}
\institute{Aalborg University, Aalborg, Denmark\\{\small \email{dave@create.aau.dk}}\\{\small \url{https://www.titanmusic.com}}}

\maketitle
\begin{abstract}
We review the notion of representing music using point sets and argue that such representations are better adapted than sequential representations for matching patterns in unvoiced, polyphonic music, such as keyboard music. 
Musical transformations such as transposition, inversion, diminution, augmentation and retrograde can be modelled by geometric transformations in pitch--time representations that combine translation with scaling parallel to and reflection in the time axis. 
We identify eight types of geometric pitch--time representation that use chromatic pitch, morphetic pitch, morph or chroma to represent pitch and either onset time or midtime to represent time. We illustrate how these types of representation allow us to characterise different types of musical transformation. For example, by using midtime instead of onset time, we can precisely characterise certain retrograde relationships; and by using morph and chroma pitch representations we can characterise transformations involving octave displacements and duplications.
We present the concept of a {\em transformation class\/} and consider the three specific classes, $F_{\mathrm{2STR}}$, $F_{\mathrm{2STRMod7}}$ and $F_{\mathrm{2STRMod12}}$. 
We introduce the notion of an {\em inter-pattern transformation graph\/} for a set of patterns, $S$, and a transformation class, $F$. Each vertex in such a graph represents a pattern in $S$ and there is an edge in the graph from pattern $P_1$ to pattern $P_2$ if and only if $P_1$ can be mapped onto $P_2$ by a transformation in $F$.
We show, with the aid of such graphs, that the musical relationships between the  occurrences of the HAYDN theme in Ravel's {\em Menuet sur le nom d'Haydn\/} can be precisely described in terms of transformations in $F_{\mathrm{2STR}}$, $F_{\mathrm{2STRMod7}}$ and $F_{\mathrm{2STRMod12}}$ within the pitch--time representations considered.
\end{abstract}
%
%
%

\section{Introduction}
\label{chapter01-introduction}

The main purpose of this paper is to review the idea of representing music using point sets and to present a number of specific types of point-set representation that prove to be particularly useful when one is interested in finding occurrences of patterns in music. Specifically, we consider the types of geometric representation that we need in order to 
find {\em transformed\/} matches of a user-supplied musical {\em query pattern\/} in a symbolic music encoding that we call a {\em dataset} that typically represents a piece or a collection of pieces. We assume that both the query pattern and the dataset are {\em polyphonic}. We also assume that they are {\em unvoiced}---that is, that the encodings of the query pattern and dataset do not contain explicit information about the voice or part to which each note belongs.

This problem of transformed pattern matching in unvoiced, polyphonic, symbolic music encodings can arise when the dataset is a symbolic encoding of a score (e.g., in Lilypond, MusicXML or MEI format) or a performance on a digital instrument (e.g., MIDI data), where information about voice structure is either entirely absent or needs to be ignored because it is unreliable, ambiguous or subjective. The problem is also relevant in the case of music which {\em is\/} unambiguously structured into voices (e.g., a baroque keyboard fugue, a string quartet or a choral work), but where one is interested in finding matches of a pattern that may contain notes from more than one voice.

\begin{figure}[t]
\centering
\resizebox{\linewidth}{!}{\includegraphics{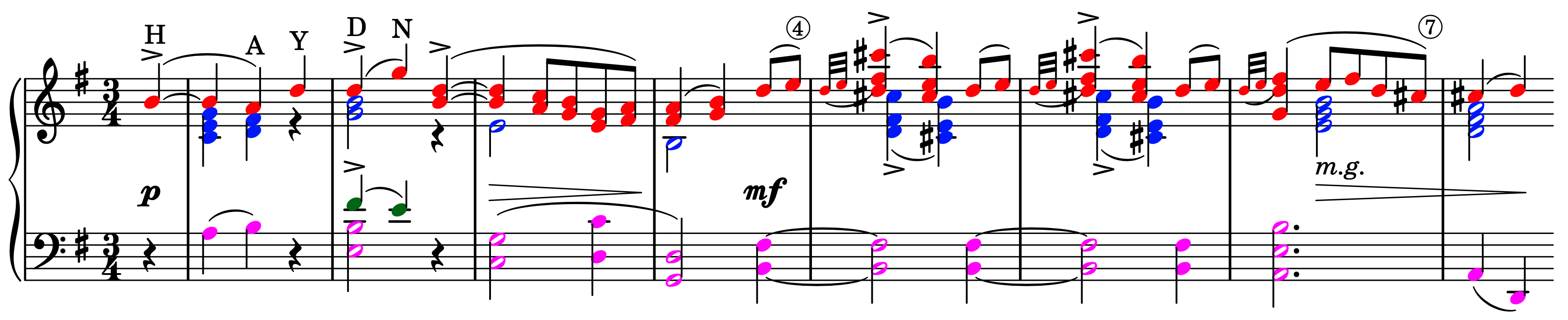}}
\caption{Bars 0 to 8 of Maurice Ravel's {\em Menuet sur le nom d'Haydn}, showing one way of partitioning the notes into four ``voices'', indicated by the colours of the noteheads.}
\label{ravel-menuet-bars1-8-voiced}
\end{figure}

\begin{figure}[t]
\centering
\resizebox{.7\linewidth}{!}{\includegraphics{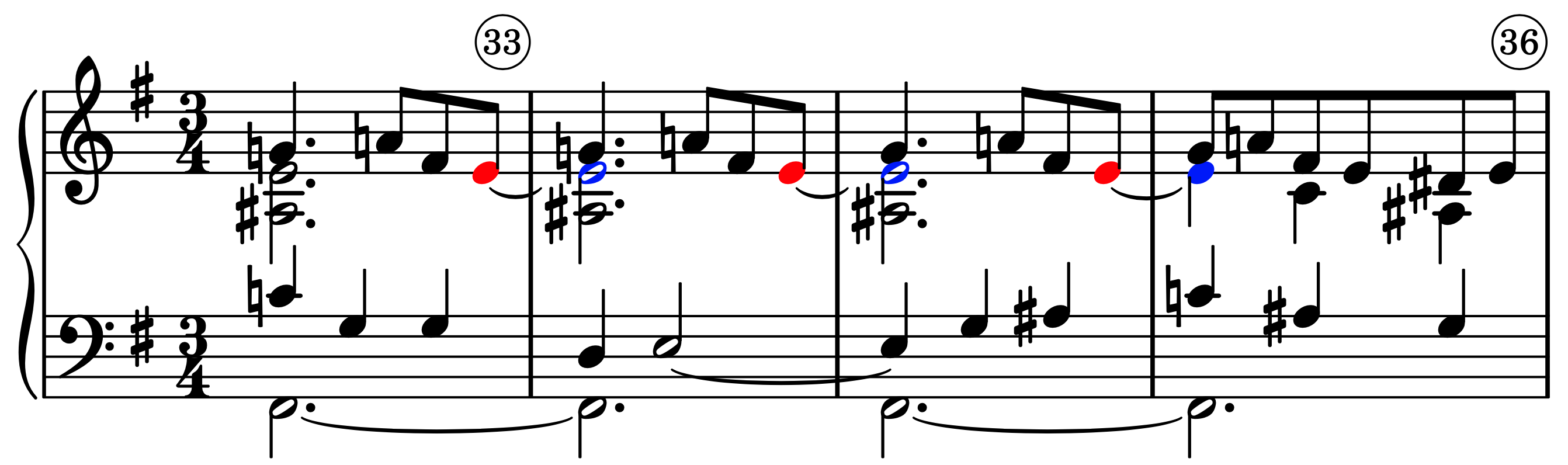}}
\caption{Bars 33 to 36 of Maurice Ravel's {\em Menuet sur le nom d'Haydn}, showing three examples of a note in one voice being tied to a note in another.}
\label{ravel-menuet-bars33-36}
\end{figure}

The problem is particularly pertinent in the context of keyboard music, where it is not uncommon for voice structure to be ambiguous, as a result of the music not being explicitly constructed from concurrent parts associated with different instruments or singing voices. 
Indeed, in keyboard music, a voice can be something quite different from a monophonic vocal or instrumental part in a choral or  orchestral work. 
A voice in keyboard music may consist of a homophonic sequence of chords or even a mixture of monophony and homophony, with chords containing a varying number of notes. Indeed, in keyboard music, a voice often corresponds more closely to an auditory stream \citep{Bregman1990}, consisting of a sequence of events that are intended to be sequentially integrated in perception, than it does to a vocal part in a choral work.
An example of this can be seen in the upper voice (coloured in red) in Fig.~\ref{ravel-menuet-bars1-8-voiced}, which shows bars 0--8 of Maurice Ravel's {\em Menuet sur le nom d'Haydn}, composed in 1909.\footnote{In this paper, we use the convention of numbering bars so that the first complete bar is assigned the number 1. Ravel's {\em Menuet\/} begins with a one-crotchet anacrusis which is therefore assigned the bar number 0.}   
As exemplified by this excerpt, the score of a keyboard piece typically only hints at how the music is intended to be understood as being constructed from voices. 
It does this through various means, such as beaming or slurring together sequences of consecutive notes (or chords) in the same voice, or by using opposing stem directions for notes in different voices on the same staff. 
For example, the simultaneous blue and red chords on the first beat of bar 5 in Fig.~\ref{ravel-menuet-bars1-8-voiced} are assigned to different voices because the notes are attached to two different stems going in opposite directions, corresponding to the fact that the two chords are intended to be played by different hands.  
Moreover, because keyboard music is played on a single instrument, composers and editors are not compelled to be consistent (or even logical) in the way they structure the notes in the score into different voices. 
For example, one can find cases where notes that are apparently intended to be understood to be in {\em different\/} voices are tied together---a physical impossibility in choral music (see Fig.~\ref{ravel-menuet-bars33-36}).

These examples illustrate that the score of a piece of keyboard music does not always unambiguously communicate a single voice structure that we can reasonably expect will correspond to the way that all listeners will interpret the music. Consequently, if one wants to use computational tools to find matches of a query pattern in such music, and one adopts an encoding system or search algorithm that assumes that (1) each piece has a single, well-specified voice structure, and (2) that each match of the query pattern is contained within a single voice, then one runs the risk of failing to find matches of the query pattern that either contain notes from more than one voice, or that are contained within a single voice according to a different interpretation of the voice structure from the one that is encoded. 

To address this problem, one might suggest considering ``all possible'' voice structures. However, it is not immediately obvious how one should define this universe of ``possible'' voice structures for any given score. One could, for example, be quite restrictive and require that every note in a voice must be one of the earliest notes following the end of the previous note in the same voice. However, as observed by \citet[p.~5]{Meredith2006b}, even with this restrictive definition, the number of different voice structures will still typically be an exponential function of the number of notes, making the problem of searching through all of them seemingly intractable.

\begin{figure}[t]
\centering
\resizebox{.4\linewidth}{!}{\includegraphics{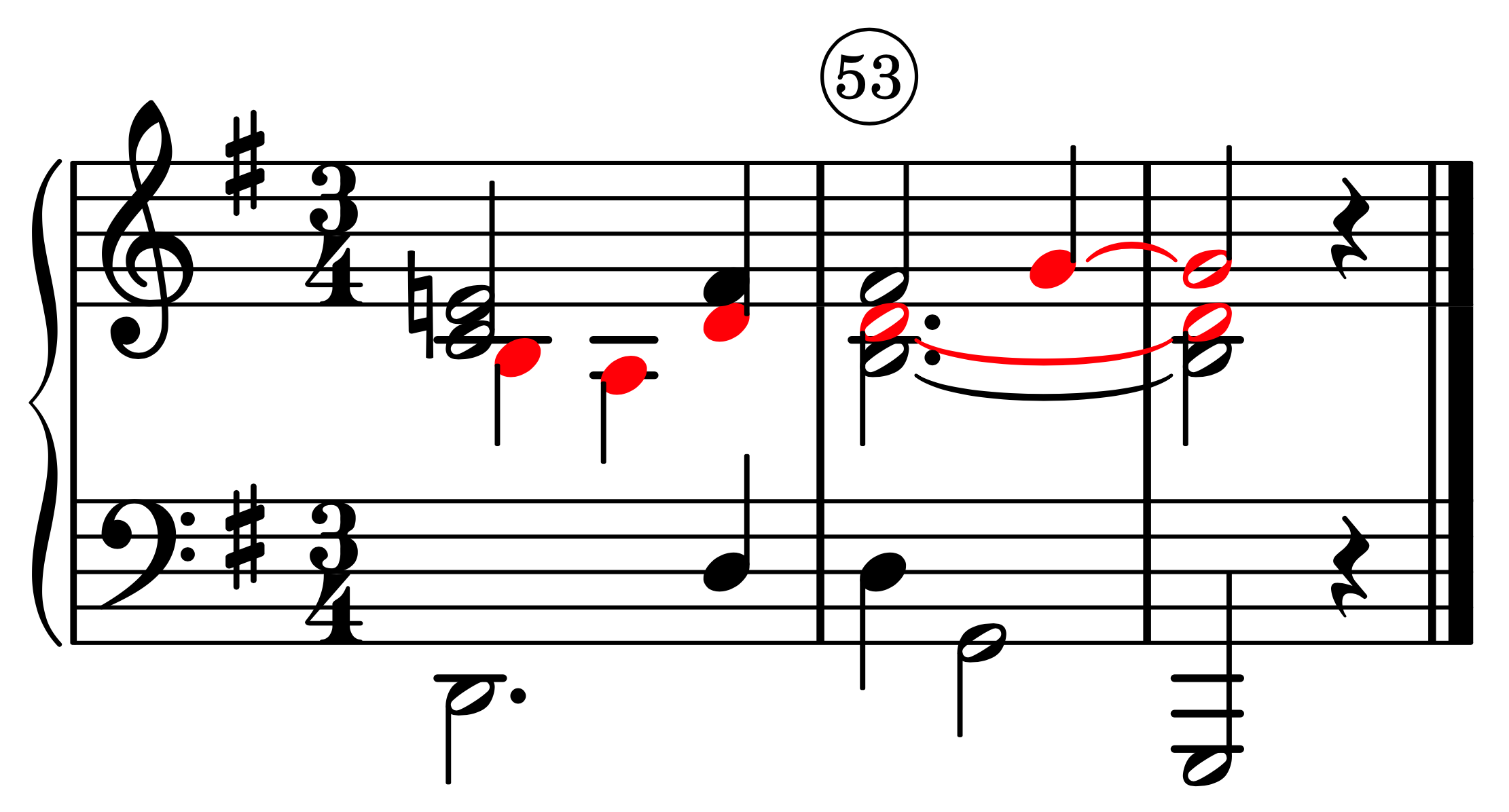}}
\caption{Bars 52 to 54 of Maurice Ravel's {\em Menuet sur le nom d'Haydn}, showing an occurrence of the HAYDN theme (in red) that includes notes that are from more than one notated ``voice'' and that overlap in time.}
\label{ravel-menuet-bars52-54}
\end{figure}

A case in point is shown in Fig.~\ref{ravel-menuet-bars52-54}, which shows an occurrence of the HAYDN theme (B--A--D--D--G) at the end of Ravel's {\em Menuet sur le nom d'Haydn\/} that seems to have been overlooked in the literature on this piece, possibly because it is obscured by the voice structure indicated by the score, according to which the occurrence contains notes from different voices. Moreover, observe that the second D4 in the theme at the beginning of bar 53 lasts beyond the onset of the final G4 that starts on the third crotchet of bar 53. If we were to adopt a strategy of searching through ``all possible'' voice structures, we would therefore, in this case, have to remove the constraint that every note within a voice must start after the end of the preceding note, increasing still further the number of ``possible'' voice structures to search. In this paper, we therefore focus on methods that do not require that the encodings used contain information about voice structure. 

\section{Occurrences of the HAYDN theme in Ravel's {\em Menuet sur le nom d'Haydn}}

Having introduced some of the challenges involved in finding matches of query patterns in unvoiced polyphonic datasets, we now consider in detail a concrete example of a symbolic music pattern-matching task that illustrates a number of important issues, such as the diversity of different ways in which a composer might transform a musical pattern, and the difficulty of deciding in certain instances even whether or not a given set of notes constitutes a ``match'' for a query pattern. This concrete problem will provide motivation for the introduction of a number of different types of point-set representation.

\begin{figure}[t]
\centering
\resizebox{.7\linewidth}{!}{\includegraphics{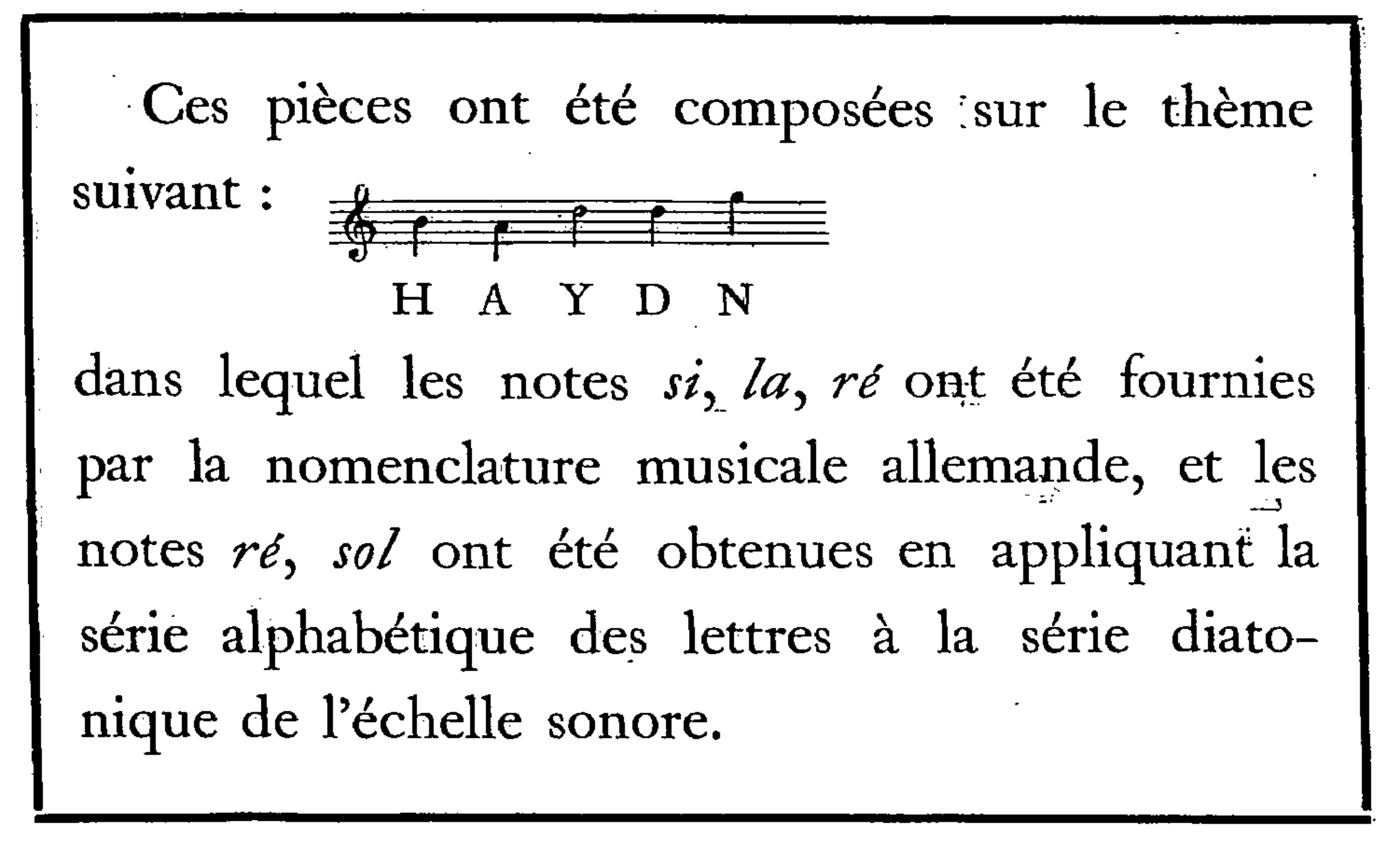}}
\caption{Explanatory note on page 2 of the 15 January 1910 issue of the {\em Revue musicale mensuelle de la Soci{\'e}t{\'e} Internationale de Musique}, preceding the collection of six pieces constituting the ``Hommage à Joseph Haydn''. Translated into English, the note reads as follows: ``These pieces have been composed on the following theme ... in which the notes B, A, D have been supplied by German musical nomenclature, and the notes D and G have been obtained by applying the sequence of letters in the alphabet to the diatonic series of the musical scale.''}
\label{haydn-hommage-explanatory-note}
\end{figure}

The problem we consider is that of finding occurrences of the HAYDN theme in the {\em Menuet sur le nom d'Haydn\/} by Maurice Ravel (1875--1937). This piece was composed in 1909 in response to a call from the {\em Revue musicale mensuelle de la Soci{\'e}t{\'e} Internationale de Musique\/} ({\em RSIM\/}) for pieces incorporating a theme encoding the word, HAYDN, on the occasion of the 100th anniversary of the death of Franz Joseph Haydn (1732--1809). Ravel's piece was one of a collection of six pieces based on the HAYDN theme, published under the title ``Hommage à Joseph Haydn" in the {\em RSIM\/} on 15 January 1910 \citep{HommageAJosephHaydn1910}, along with a number of articles relating to Haydn. The other pieces in this ``Hommage'' were composed by Claude Debussy (1862--1918), Paul Dukas (1865--1935), Vincent d'Indy (1851--1931), Reynaldo Hahn (1874--1947) and Charles-Marie Widor (1844--1937). The ``Hommage'' itself is preceded by an explanatory note, written by the editor of the {\em RSIM}, Jules Écorcheville, shown in Fig.~\ref{haydn-hommage-explanatory-note}. This note states that the pieces were ``composed on the theme'' shown in the figure. As can be seen in Fig.~\ref{haydn-hommage-explanatory-note}, the HAYDN theme itself consists simply of the sequence of five notes, B--A--D--D--G, with `H' in the German pitch-naming system corresponding to B$\natural$, and the first D in the theme representing the letter `Y' as a result of cyclically mapping the sequence of pitch letter names, A--G, onto the letters of the alphabet.

\begin{figure}
\centering
\resizebox{\linewidth}{!}{\includegraphics{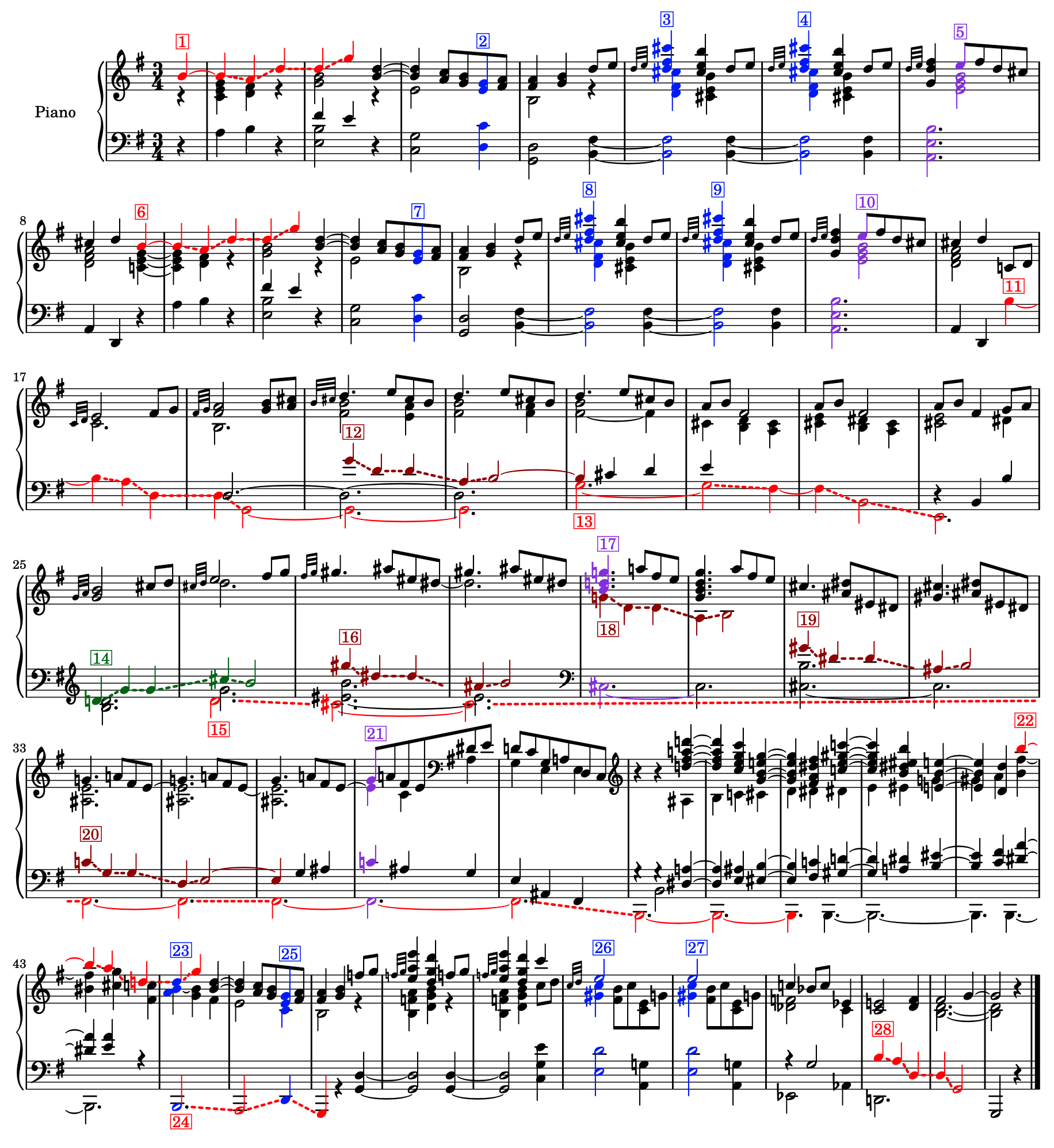}}
\caption{Occurrences of the HAYDN theme in Maurice Ravel's {\em Menuet sur le nom d'Haydn}, identified by \citet{Soucy2009}. Occurrences 1 (bars 0--2), 11 (bars 16--20), 12 (bars 19--21), 14 (bars 25--26), 22 (bars 42--44) and 28 (bars 52--53) were explicitly annotated by Ravel in the original edition of the work, published in the {\em RSIM\/} \citep{HommageAJosephHaydn1910}.}
\label{figure-soucy-haydn-occurrences}
\end{figure}

\subsection{Soucy's analysis}

\citet{Soucy2009} presents an in-depth study of the six pieces in the ``Hommage à Joseph Haydn'', including an analysis of each piece that aims to locate and characterise all the occurrences of the HAYDN theme.  
Figure~\ref{figure-soucy-haydn-occurrences} shows the 28 occurrences of the HAYDN theme in Ravel's {\em Menuet\/} identified by \citet[pp.~103--104]{Soucy2009}. Of these, occurrences 1 (bars 0--2), 11 (bars 16--20), 12 (bars 19--21), 14 (bars 25--26), 22 (bars 42--44) and 28 (bars 52--53) were explicitly annotated by Ravel in the original edition of the work, published in the {\em RSIM\/} \citep{HommageAJosephHaydn1910}. The occurrences printed in 
red 
in Fig.~\ref{figure-soucy-haydn-occurrences} are those where the notes of the theme occur in the original order, as shown in Fig.~\ref{haydn-hommage-explanatory-note} (i.e., occurrences 1, 6, 11, 13, 15, 22, 24 and 28).  Occurrences printed in 
brown (i.e., occurrences 12, 16, 18, 19 and 20)
are those in which the pitches in the theme appear in reverse (retrograde). Occurrence 14, printed in 
green,
is a retrograde inversion of the theme. The occurrences printed in 
blue
in Fig.~\ref{figure-soucy-haydn-occurrences} (and labelled `$\alpha$' in Soucy's thesis) result from ``[p]utting all of HAYDN's notes in a chord'' to produce ``a ninth chord without its seventh'' \citep[p.~67]{Soucy2009}---i.e., an added-ninth chord (or, equivalently, an added-second chord). Conversely, the occurrences printed in 
lilac
(labelled ``$1/\alpha$'' by Soucy) are those in which the notes in the {\em inversion\/} of the theme are all presented at once to form an added-fourth chord. For example, occurrence 5 consists of the notes, $\left\lbrace{\mathrm{E, G, A, B}}\right\rbrace$.

None of the 28 occurrences identified by Soucy are identical in terms of pitch and rhythm to the theme proposed by Écorcheville (i.e., the theme shown in Fig.~\ref{haydn-hommage-explanatory-note}). Thus, if one were to adopt a rather strict interpretation of what it means for a pattern to ``occur'' in a piece of music, then one might claim that Écorcheville's theme does not occur in the piece at all! However, the fact that the theme is written as a sequence of crotchets with no time signature might indicate that Écorcheville's notation was not intended to specify any particular rhythm, but only the sequence of pitch names. Occurrences 1 and 6 are the most similar to the provided theme, but even these transform it by lengthening the initial note from a crotchet to two tied crotchets. In occurrences 11, 13, 15 and 28, the rising fourths in the original version of the theme are replaced with falling fifths (or, in occurrence 15, a falling twelfth). In occurrence 22 only the first of the rising fourths in the original theme (from `A' to `Y') is replaced with a falling fifth, whereas, in occurrence 24, it is only the second rising fourth in the original theme (from `D' to `N') that is transformed into a falling fifth. The note durations in occurrence 1 are repeated exactly in occurrences  6, 14, 16, 18, 19 and 22. In occurrences 11, 12 and 20, the durations are the same as in 1, except that the last note in the occurrence is lengthened. 
In occurrences 13, 15 and 24, the duration of every note is longer than that of its counterpart in occurrence 1. However, these three occurrences are not strict augmentations of 1, as the scale factor by which durations are multiplied does not remain constant within each occurrence. Occurrence 15 is remarkable due to the fact that there is a rest of four bars between its second and third tones.\footnote{Strictly speaking, a {\em tone\/} in psychoacoustics is defined to be a single sound that has a pitch \citep[pp.~3, 404]{Moore2008}. Here, we consider a sequence of tied notes in a score to be a single `tone', since, when performed, such a sequence has a single onset and typically results in a single tone in the psychoacoustic sense. A single untied note also typically results in a single tone when performed. Thus, in occurrence 1, for example, the second {\em note\/} is the second B4, at the beginning of bar 1 that is tied to the opening B4; but the second {\em tone\/} in occurrence 1 is the A4 on the second beat of bar 1.\label{footnote-note-and-tone}}

Looking more closely at the cases where the notes of the theme are expressed simultaneously as a chord, we observe that occurrences 2, 7 and 25 form C major added-ninth chords (although they function as dominant 11th chords). Occurrences 3, 4, 8 and 9 form B minor added-ninth chords, while occurrences 26 and 27 consist of augmented triads on C with an added ninth or second (D4). Occurrences 5 and 10 form E minor added fourth chords, while occurrence 17 is a G major triad with an added (sharpened) fourth in the bass, and occurrence 21 is a C major triad, also with an added sharpened fourth (again in the bass). Occurrence 23 seems to be an error, since this chord lacks the G that would be required to form an added second chord. Perhaps Soucy misread the bass note here as being on a treble staff.

Occurrences 13 and 24 each lack a note corresponding to the second D in the theme: in 13, we would expect there to be two consecutive Bs and in 24 we would expect two consecutive Ds. These two occurrences should therefore be considered {\em partial\/} matches of Écorcheville's theme. 

In occurrence 15, there is a particularly long temporal gap between the C$\sharp$4 in bars 27--28 and the F$\sharp$2 that starts in bar 33. One could argue that the C$\sharp$3s in bars 29--32 provide a kind of ``pitch class continuity'' from the C$\sharp$4 in bars 27--28, while also providing registral ``stepping stones'' down to the F$\sharp$2 in bar 33. On this interpretation, this occurrence should perhaps be considered to include the C$\sharp$3s in bars 29--32, but this would imply a transformation of Écorcheville's theme that involves repeating the second note in the theme multiple times. If we are to allow for the pattern to be transformed in such a way, then one could then also argue that this occurrence of the theme should be considered to include the further repetitions of the B1 in bars 40--43. This example illustrates some of the difficulties that can arise when trying to decide whether a particular pattern of notes should be considered a ``match'' or ``occurrence'' of some given query pattern.\label{what-is-an-occurrence} Sometimes it is not at all obvious exactly which notes should be included in and excluded from such a match. In the case of occurrence 15, should we or should we not include the trailing B1s in bars 40--43 or the ``stepping-stone'' C$\sharp$3s  in bars 29--32? 

In any case, according to Soucy, Ravel's {\em Menuet\/} contains 14 melodic statements of the HAYDN theme (occurrences 1, 6, 11, 12, 13, 14, 15, 16, 18, 19, 20, 22, 24 and 28), of which five are in retrograde (occurrences 12, 16, 18, 19, and 20), and one is a retrograde inversion (occurrence 14). In addition, according to Soucy, the piece contains 10 occurrences of the theme in which the notes of the theme are played simultaneously to form an added-ninth (or added-second) chord (occurrences 2, 3, 4, 7, 8, 9, 23, 25, 26 and 27). However, as discussed above, occurrence 23 has to be considered only a partial occurrence. The piece also contains, according to Soucy, four occurrences in which notes with the pitch letter names of a transposed inversion of the theme are played simultaneously to form added-fourth chords (occurrences 5, 10, 17 and 21).  

\begin{figure}[t]
	\centering
	\subfloat[Added-ninth chords from the end of the first movement of Ravel's {\em Sonatine}, composed in 1903--05.]{\resizebox{.5\textwidth}{!}{\includegraphics{"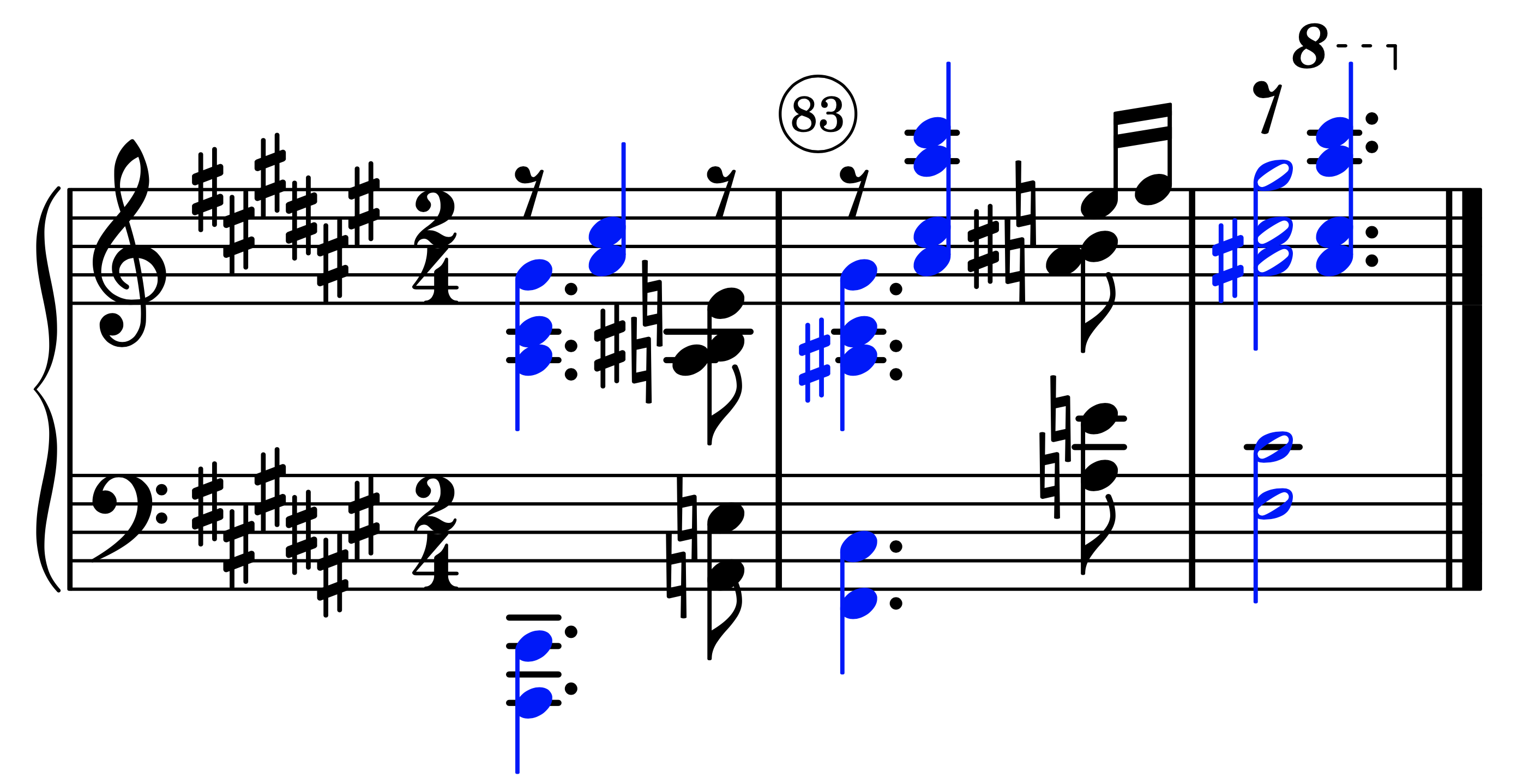"}}}\\
	\subfloat[Added-fourth chords in bar 29 of Ravel's {\em Jeux d'eaux}, composed in 1901.]{\resizebox{.6\textwidth}{!}{\includegraphics{"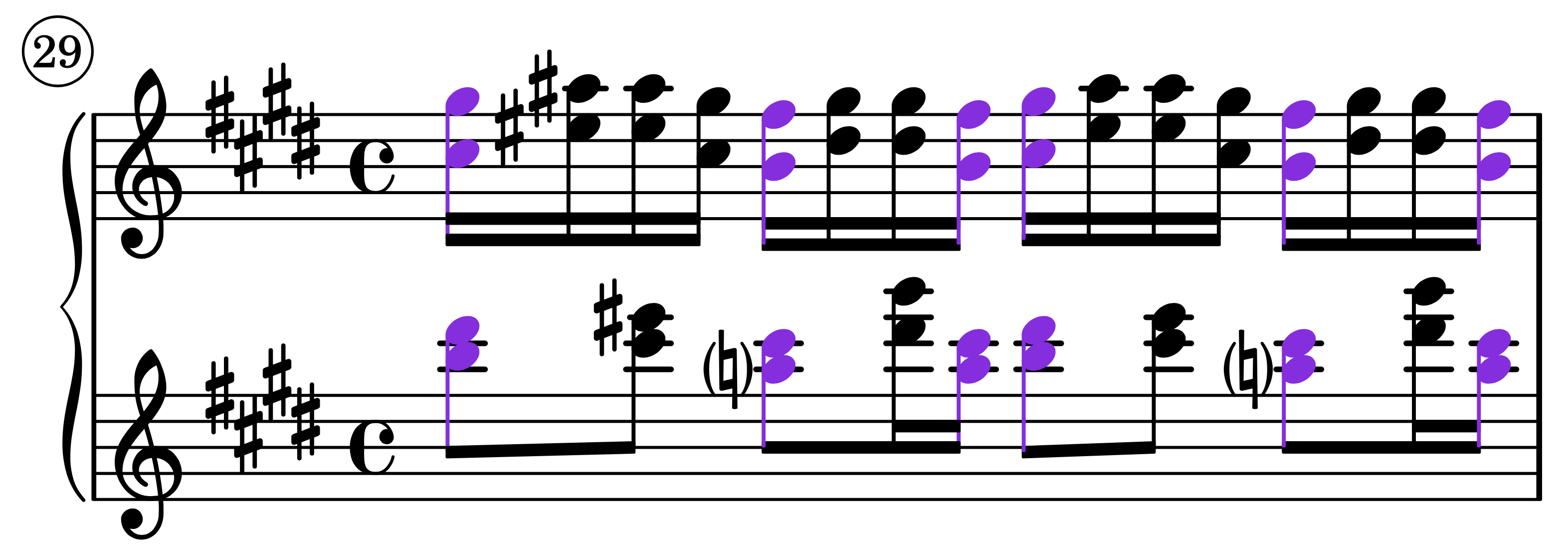"}}}
	\caption{Examples of added-fourth chords (in lilac) and added-ninth chords (in blue) from other works by Maurice Ravel, suggesting that such chords in the {\em Menuet sur le nom d'Haydn\/} may not have been intended to be occurrences of the HAYDN theme.}
	\label{added-ninth-fourth-chords-in-ravel}
\end{figure}


Added ninth (or second) chords and added fourth (or eleventh) chords occur frequently throughout Ravel's œuvre, as \citet[p.~21]{Soucy2009} himself acknowledges. Figure~\ref{added-ninth-fourth-chords-in-ravel} shows some examples of added fourth and added ninth chords in other pieces by Ravel. One might therefore question whether Ravel deliberately intended such chords in the {\em Menuet\/} to be interpreted as references to the HAYDN theme. 

\begin{figure}[t]
	\centering
	\subfloat[]{\resizebox{!}{2cm}{\includegraphics{"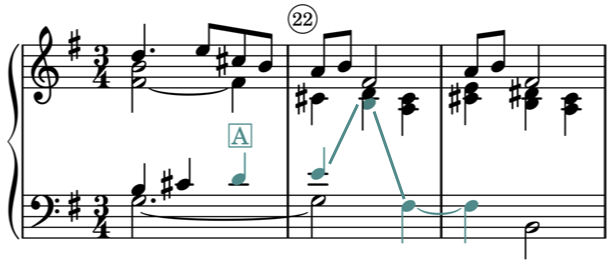"}}}\\
	\subfloat[]{\resizebox{!}{2cm}{\includegraphics{"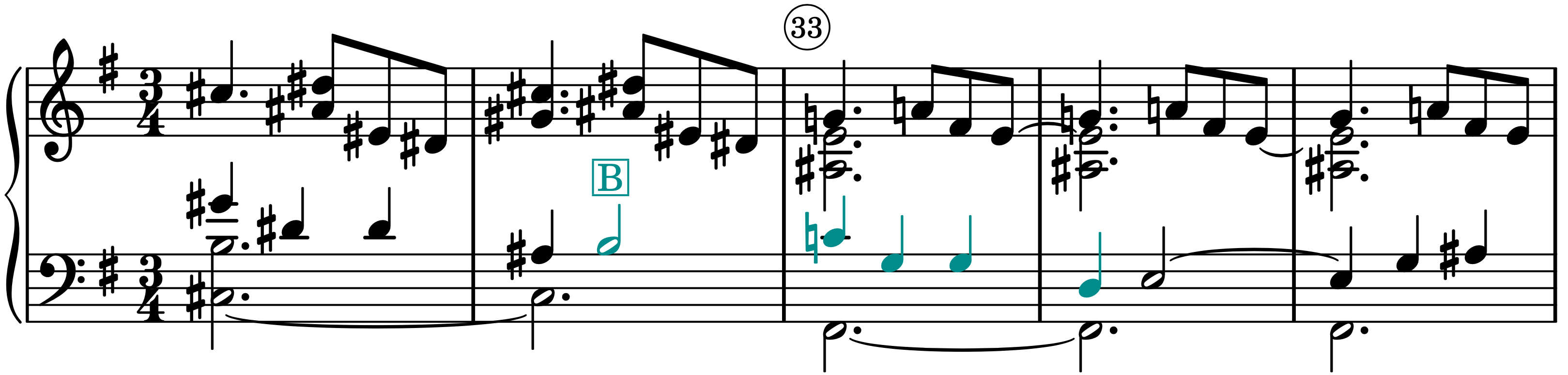"}}}\\
	\subfloat[]{\resizebox{!}{2cm}{\includegraphics{"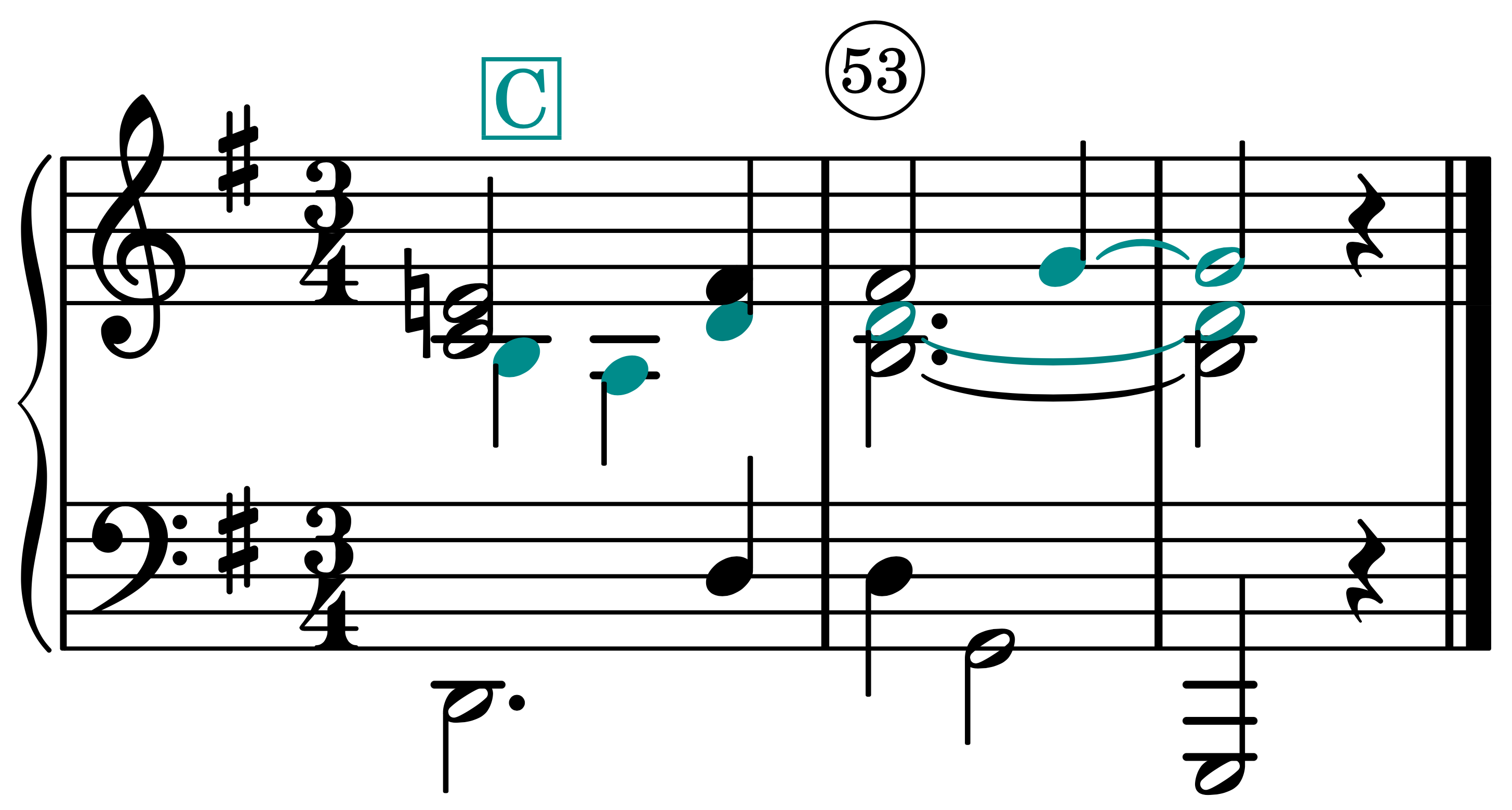"}}}
	\caption{Three further occurrences of the HAYDN theme in the {\em Menuet}, that were not identified by either Ravel or Soucy: (a) a partial, inverted occurrence of the HAYDN theme in bars 21--23; (b) an inversion of the HAYDN theme at bars 32--34; and (c) a statement of the HAYDN theme at bars 52--54 that includes notes from different voices, some of which overlap in time.}
	\label{further-occurrences-of-HAYDN-theme}
\end{figure}


\subsection{Further occurrences of the HAYDN theme in Ravel's {\em Menuet}}

Figure~\ref{further-occurrences-of-HAYDN-theme} shows three further occurrences of the HAYDN theme in Ravel's {\em Menuet}, that appear to have been overlooked in previous writings about this piece. These are labelled A, B and C in Fig.~\ref{further-occurrences-of-HAYDN-theme}. Occurrence A in bars 21--23 is a partial, inverted occurrence of the theme where one of the Ds in the original theme is not represented (as is the case in occurrences 13 and 24 in Fig.~\ref{figure-soucy-haydn-occurrences}). Occurrence B (bars 32--34) is a complete inversion of the theme that interestingly intersects with occurrences 19 and 20 and serves to ``bind'' them together. Finally, occurrence C (bars 52--54) is the occurrence identified above in Fig.~\ref{ravel-menuet-bars52-54} that interestingly includes notes from different notated ``voices'', some of which overlap in time.



\section{Geometric representations and transformations}

In this section, we identify a number of different ways of representing a musical ``object''---typically a passage, a piece or a collection of pieces---as a set of geometric points. We motivate each type of representation by showing how it is well adapted for discovering occurrences of a query pattern in a dataset under different classes of geometric transformation that correspond to commonly-used classes of musical transformation.
%
%
We assume that both the query pattern and the dataset are encodings of sets of notes or tones (as defined in footnote~\ref{footnote-note-and-tone}) rather than, say, sequences of audio samples. For example, the query pattern and dataset encodings might have been derived from notation by optical music recognition or by encoding a score manually or with music typesetting software, such as LilyPond\footnote{\url{https://lilypond.org/}}, MuseScore\footnote{\url{https://musescore.org/en}} or Verovio\footnote{\url{https://www.verovio.org/index.xhtml}}. Alternatively, the query pattern and dataset might be derived from MIDI files created by performing a piece on a digital instrument. More specifically, it is assumed that the query pattern and the dataset encode at least the {\em onset time\/} and the {\em chromatic pitch\/} \citep[p.~127]{Meredith2006} of each tone in the score. Some of the representation types introduced also require that we know the {\em duration\/} of each tone.

\subsection{Chromatic-pitch--onset (CPO) representation}
As explained by \citet[p.~127]{Meredith2006}, the {\em chromatic pitch\/} of a note is an integer that corresponds to the key on a normal piano keyboard that would have to be pressed in order to play the note. Raising the pitch of the note by one semitone increases the chromatic pitch by 1. We define the chromatic pitch of A0 to be 0. The chromatic pitch of middle C (C4) is therefore 39 and the chromatic pitch of concert A (A4) is 48. Enharmonically equivalent pitch names have the same chromatic pitch as each other---for example, B$\sharp$3, C4 and D$\doubleflat$4 all have a chromatic pitch of 39. For notes with pitch names that have MIDI note numbers (i.e., ones that are enharmonically equivalent to pitch names from C$-1$ to G9), the chromatic pitch is 21 less than the MIDI note number. Note that, unlike a MIDI note number, a chromatic pitch can, in principle, take any integer value---positive, zero or negative.

\begin{sidewaysfigure}
\resizebox{1\linewidth}{!}{\includegraphics{"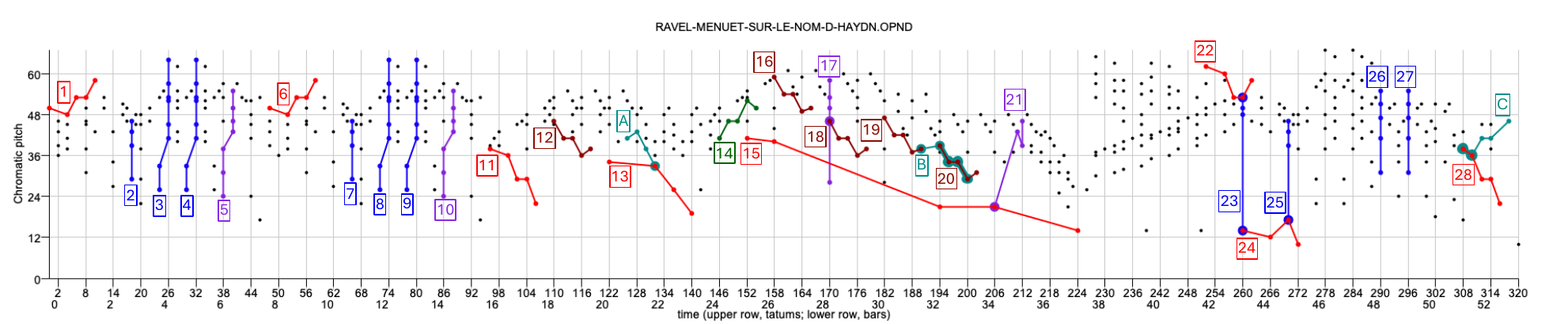"}}
\caption{A {\em chromatic-pitch--onset\/} (or CPO) representation of Ravel's {\em Menuet sur le nom d'Haydn}. Each point represents a distinct untied note or sequence of tied notes in the score. The $x$-coordinate of a point gives the onset time of the corresponding note in tatums (in this case, quavers) and the $y$--coordinate gives the chromatic pitch of the note. The occurrences of the HAYDN theme identified in Figs.~\ref{figure-soucy-haydn-occurrences} and \ref{further-occurrences-of-HAYDN-theme} are shown, labelled with the same numbers and letters and coloured using the same colours as in these previous figures. When a point occurs in two patterns, it is drawn twice with different diameters and colours so that its presence in both patterns is visible.}
	\label{RAVEL-HAYDN-GROUND-TRUTH-CPO-GT}
\end{sidewaysfigure}

Figure~\ref{RAVEL-HAYDN-GROUND-TRUTH-CPO-GT} shows a geometric representation of Ravel's {\em Menuet sur le nom d'Haydn\/} as a set of 2-dimensional points in which each point, $\langle t,p\rangle$, represents a tone in the score (i.e., either an untied note or a sequence of tied notes, as defined in footnote~\ref{footnote-note-and-tone}), where $t$ is the onset time of the tone in tatums (in this case, quavers) and $p$ is the chromatic pitch of the note. We call such a geometric, point-set representation a {\em chromatic-pitch--onset\/} (or CPO) representation. The 28 occurrences of the HAYDN theme identified by Soucy shown in Fig.~\ref{figure-soucy-haydn-occurrences} are also indicated in Fig.~\ref{RAVEL-HAYDN-GROUND-TRUTH-CPO-GT}, along with the three further occurrences shown in Fig.~\ref{further-occurrences-of-HAYDN-theme}.

As can be seen in Fig.~\ref{RAVEL-HAYDN-GROUND-TRUTH-CPO-GT}, some of the occurrences of the HAYDN theme in the {\em Menuet\/} have representations within the CPO representation that are related to each other by simple geometric transformations. For example, within this representation, occurrences 12, 18 and 20 are related to each other by translation: occurrences 18 and 20 can be obtained by translating occurrence 12 by the vectors $\langle60,0\rangle$ and $\langle84,-7\rangle$, respectively. We can say that occurrences 18 and 20 are {\em complete, exact matches\/} of occurrence 12 within the CPO representation under the geometric transformation of translation. Indeed,  within this representation, occurrences 12, 18 and 20 are exact, complete matches of {\em each other\/} under translation---so we can say that they are {\em translationally equivalent\/} within this representation. Similarly, within this CPO representation, each of the following sets of occurrences forms a set of translationally equivalent patterns: $\lbrace1,6\rbrace$, $\lbrace2,7\rbrace$, $\lbrace3,4,8,9\rbrace$, $\lbrace5,10\rbrace$, $\lbrace16,19\rbrace$ and $\lbrace26,27\rbrace$. 

\subsection{Retrogrades and the chromatic-pitch--midtime (CPM) representation}
\begin{sidewaysfigure}
\resizebox{\linewidth}{!}{\includegraphics{"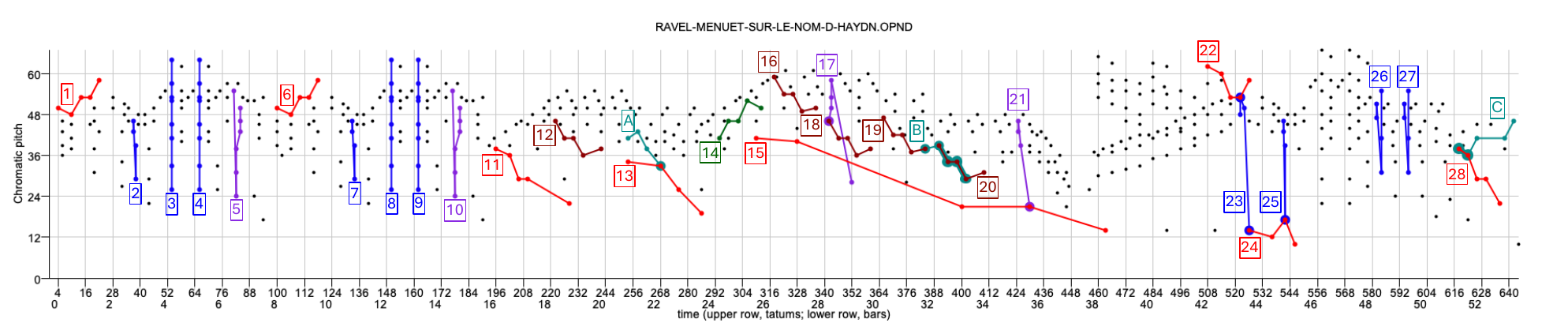"}}
\caption{A {\em chromatic-pitch--midtime\/} or CPM representation of Ravel's {\em Menuet sur le nom d'Haydn}. Each point represents a distinct untied note or sequence of tied notes in the score. The $x$-coordinate of a point gives the midtime of the corresponding note in tatums (in this case, semiquavers) and the $y$--coordinate gives the chromatic pitch of the note. The occurrences of the HAYDN theme identified in Figs.~\ref{figure-soucy-haydn-occurrences} and \ref{further-occurrences-of-HAYDN-theme} are shown, labelled with the same numbers and letters and coloured using the same colours as in these previous figures. When a point occurs in two patterns, it is drawn twice with different diameters and colours so that its presence in both patterns is visible.}
	\label{RAVEL-HAYDN-GROUND-TRUTH-CPM-GT}
\end{sidewaysfigure}

Let us now consider the relationship between occurrence 1 and occurrence 18 in Fig.~\ref{figure-soucy-haydn-occurrences}. We see from the score in Fig.~\ref{figure-soucy-haydn-occurrences} that occurrence 18 is an exact transposed retrograde of occurrence 1 that preserves the tone durations and the chromatic pitch intervals between the tones. However, occurrence 18 is {\em not\/} an exact retrograde of occurrence 1 in the CPO representation in Fig.~\ref{RAVEL-HAYDN-GROUND-TRUTH-CPO-GT}, because the $x$-axis in this representation represents {\em onset\/} time, and it is the {\em offset\/} of the first tone in occurrence 1 that is mapped onto the {\em onset\/} of the last tone in occurrence 18. Clearly, we would like the exact musical retrograde of a pattern to be related to the pattern by a reflection in the pitch axis followed by a translation in pitch--time space. If we know the duration of each tone in the dataset and query, then we can achieve this by using a different 2-dimensional point-set representation where the $x$-coordinate represents the instant in time that is {\em exactly half-way through\/} the duration of the tone. We will use the term, {\em mid-duration time\/} or {\em midtime\/} to refer to this property of a tone. In an exact retrograde, the midtimes of the tones in a pattern map onto the midtimes of the corresponding tones in the retrograde occurrence of the pattern. Figure~\ref{RAVEL-HAYDN-GROUND-TRUTH-CPM-GT} shows such a representation of Ravel's {\em Menuet}. We refer to this type of representation as a {\em chromatic-pitch--midtime\/} (or CPM) representation. Note that, in this CPM representation, occurrence 1 is mapped exactly onto occurrence 18 by multiplying the midtimes (i.e., the $x$-coordinates of its points) by a factor of $-1$ (i.e., reflecting in the pitch axis, $x=0$) and then translating by $\langle364,-12\rangle$. On the other hand, the set of occurrences, $\lbrace12,18,20\rbrace$, that were translationally equivalent in the CPO representation in Fig.~\ref{RAVEL-HAYDN-GROUND-TRUTH-CPO-GT} are {\em not\/} translationally equivalent in the CPM representation in Fig.~\ref{RAVEL-HAYDN-GROUND-TRUTH-CPM-GT}, because occurrence 18 is no longer translationally equivalent to occurrences 12 and 20. This example suggests that, if we are interested in finding retrogrades (or retrograde inversions) of a query pattern, then we should search in a CPM representation like the one in Fig.~\ref{RAVEL-HAYDN-GROUND-TRUTH-CPM-GT}, where the $x$-coordinate of each point represents the midtime of each tone rather than its onset. However, we need to be aware that patterns that are translationally equivalent in the CPO representation may {\em not\/} be translationally equivalent in the CPM representation.

\subsection{Morphetic-pitch--time representations}
Let us next consider the geometric relationships between occurrence 1 and occurrences 14, 16, 19 and B. Occurrences 16 and 19 fail to be exact transposed retrogrades of occurrence 1 because the pitch interval between the fourth and fifth tones in each of these occurrences is a rising {\em minor\/} second, whereas the corresponding pitch interval between the first and second tones in occurrence 1 is a falling {\em major\/} second. Similarly, occurrence 14 fails to be an exact transposed retrograde inversion of occurrence 1 because the interval between its third and fourth tones is a rising {\em augmented\/} fourth, whereas the interval between the second and third tones in occurrence 1 is a rising {\em perfect\/} fourth. Occurrence B fails to be an exact, transposed inversion of occurrence 1 because the interval between its first two tones is a rising {\em minor\/} second, whereas the interval between the first two tones of occurrence 1 is a falling {\em major\/} second. This means that an algorithm that is only capable of finding exact complete matches of a point pattern when it is translated and also possibly reflected in either the pitch or the time axis in a pitch--time representation would be incapable of identifying occurrences 14, 16, 19 or B as matches of occurrence 1 in either the CPO or CPM representations.

\begin{sidewaysfigure}
\resizebox{\linewidth}{!}{\includegraphics{"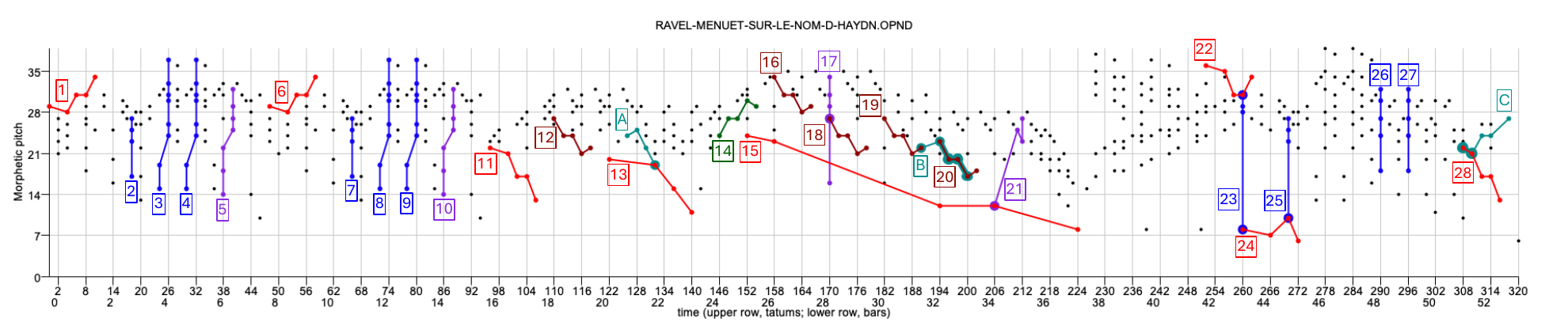"}}
\caption{A {\em morphetic pitch--onset\/} (or MPO) representation of Ravel's {\em Menuet sur le nom d'Haydn}. Each point represents a distinct untied note or sequence of tied notes in the score. The $x$-coordinate of a point gives the onset of the corresponding note in tatums (in this case, quavers) and the $y$--coordinate gives the morphetic pitch of the note. The occurrences of the HAYDN theme identified in Figs.~\ref{figure-soucy-haydn-occurrences} and \ref{further-occurrences-of-HAYDN-theme} are shown, labelled with the same numbers and letters and coloured using the same colours as in these previous figures. When a point occurs in two patterns, it is drawn twice with different diameters and colours so that its presence in both patterns is visible.}
	\label{RAVEL-HAYDN-GROUND-TRUTH-MPO-GT}
\end{sidewaysfigure}

\begin{sidewaysfigure}
\resizebox{\linewidth}{!}{\includegraphics{"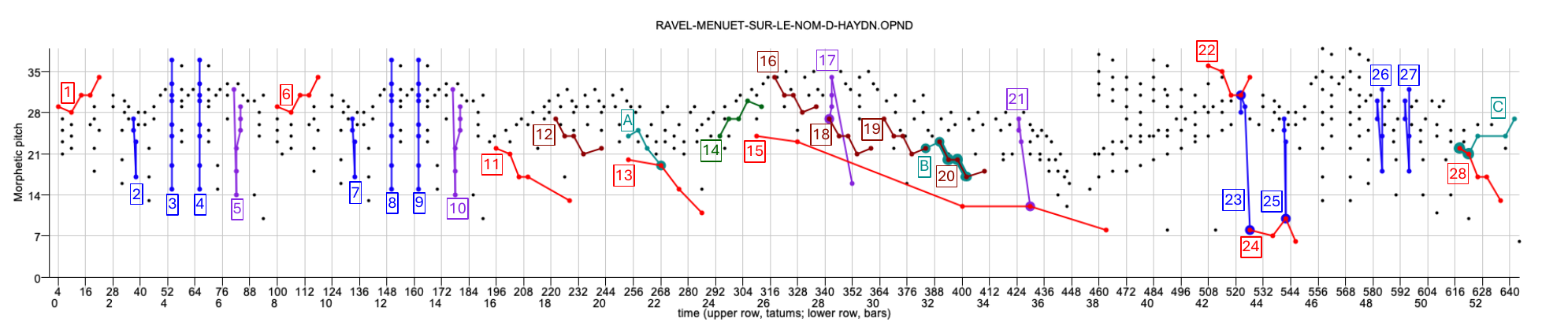"}}
\caption{A {\em morphetic-pitch--midtime\/} (or MPM) representation of Ravel's {\em Menuet sur le nom d'Haydn}. Each point represents a distinct untied note or sequence of tied notes in the score. The $x$-coordinate of a point gives the midtime of the corresponding note in tatums (in this case, semiquavers) and the $y$--coordinate gives the morphetic pitch of the note. The occurrences of the HAYDN theme identified in Figs.~\ref{figure-soucy-haydn-occurrences} and \ref{further-occurrences-of-HAYDN-theme} are shown, labelled with the same numbers and letters and coloured using the same colours as in these previous figures. When a point occurs in two patterns, it is drawn twice with different diameters and colours so that its presence in both patterns is visible.}
	\label{RAVEL-HAYDN-GROUND-TRUTH-MPM-GT}
\end{sidewaysfigure}

We can solve this problem by replacing chromatic pitch in the CPO and CPM representations with {\em morphetic pitch}. As explained by \citet[p.~127]{Meredith2006}, the {\em morphetic pitch\/} of a note is an integer that is determined by 
\begin{enumerate}
\item the vertical position of the note-head on the staff,
\item the clef in operation on the staff at the location of the note, and
\item the transposition of the staff.
\end{enumerate}
The morphetic pitch of a note is independent of the sounding pitch of a note and the chromatic pitch of the note. Moving a note up one step on a staff (while keeping the clef and transposition of the staff constant) increases its morphetic pitch by 1. The morphetic pitch of A0 is defined to be 0. The morphetic pitch associated with a pitch name is independent of its accidental, so A0, A$\flat$0, A$\sharp$0 and A$\doublesharp$0 all have a morphetic pitch of 0. The morphetic pitch of middle C (C4) is 23. Given two pitch names, $N_1$ and $N_2$, it is possible for the chromatic pitch of $N_1$ to be greater than that of $N_2$ while the morphetic pitch of $N_2$ is greater than that of $N_1$. This occurs if, for example, $N_1$ is B$\sharp$3 and $N_2$ is C$\flat$4. If the input encodings we use for the dataset and query pattern provide the pitch name of each note, then we can compute both the chromatic pitch and the morphetic pitch of each note directly from its pitch name using the P2PN algorithm provided by \citet[p.~43]{Meredith2007b}. If an input encoding does not provide information about the pitch name of each note but does provide the onset time and chromatic pitch (or MIDI note number) of each note, then the morphetic pitch of each note can be estimated using a pitch spelling algorithm such as the PS13s1 algorithm described by \citet[pp.~143--145]{Meredith2006}.

Figure~\ref{RAVEL-HAYDN-GROUND-TRUTH-MPO-GT} shows a 2-dimensional point set representing Ravel's {\em Menuet}, in which each point, $\langle t,p\rangle$, represents either an untied note or a sequence of tied notes in the score, and where $t$ is the onset time of the note and $p$ is the morphetic pitch of the note. We refer to this type of representation as a {\em morphetic-pitch--onset\/} (or MPO) representation. Observe that, in this MPO representation, occurrences 12, 16, 18, 19 and 20 are translationally equivalent and that occurrence 1 can be mapped exactly onto occurrence B by translating by the vector $\langle190,-51\rangle$ and then reflecting in the $x$-axis. Moreover, occurrence 14 can be mapped exactly onto occurrences 12, 16, 18, 19 and 20 simply by translating by the appropriate vector and then reflecting in the $x$-axis. However, because the time axis indicates {\em onset\/} time, we still do not have exact matches in this MPO representation under translation combined with reflections in the $x$- and/or $y$-axes between occurrence 1 and occurrences 14, 16, 18 and 19. To achieve this, we need to use a {\em morphetic-pitch--midtime\/} (or MPM) representation, as shown in Fig.~\ref{RAVEL-HAYDN-GROUND-TRUTH-MPM-GT}. Observe that, in this MPM representation, occurrence 14 can be obtained exactly from occurrence 1 by reflecting it first in the $y$-axis (retrograde), then translating by $\langle316,-58\rangle$, and then reflecting in the $x$-axis. Occurrence 1 can be mapped exactly onto occurrences 16, 18 and 19 in the MPM representation by first reflecting in the $y$-axis and then translating by $\langle340,0\rangle$, $\langle364, -7\rangle$ and $\langle388,-7\rangle$, respectively.

\begin{sidewaystable}
\caption{Inter-pattern transformation graph for the set of patterns in the CPO representation in Fig.~\ref{RAVEL-HAYDN-GROUND-TRUTH-CPO-GT} and the transformation class, $F_{\mathrm{2STR}}$.}
\label{RAVEL-HAYDN-GROUND-TRUTH-CPO-IPTG}
\resizebox{\linewidth}{!}{
\begin{tabularx}{1.33\linewidth}{ll|lllllllllllllllllllllllllllllll}
&&\multicolumn{31}{c}{{\bfseries To pattern}}\\
&&1&2&3&4&5&6&7&8&9&10&11&12&13&14&15&16&17&18&19&20&21&22&23&24&25&26&27&28&A&B&C\\\cline{1-33}
&1 &$I$ & & & & &$T_{\langle48,0\rangle}$ & & & & & & & & & & & & & & & & & & & & & & & & & \\\cline{2-33}
&2 & &$I$ & & & & &$T_{\langle48,0\rangle}$ & & & & & & & & & & & & & & & & & & & & & & & & \\\cline{2-33}
&3 & & &$I$ &$T_{\langle6,0\rangle}$ & & & &$T_{\langle48,0\rangle}$ &$T_{\langle54,0\rangle}$ & & & & & & & & & & & & & & & & & & & & & & \\\cline{2-33}
&4 & & &$T_{\langle-6,0\rangle}$ &$I$ & & & &$T_{\langle42,0\rangle}$ &$T_{\langle48,0\rangle}$ & & & & & & & & & & & & & & & & & & & & & & \\\cline{2-33}
&5 & & & & &$I$ & & & & &$T_{\langle48,0\rangle}$ & & & & & & & & & & & & & & & & & & & & & \\\cline{2-33}
&6 &$T_{\langle-48,0\rangle}$ & & & & &$I$ & & & & & & & & & & & & & & & & & & & & & & & & & \\\cline{2-33}
&7 & &$T_{\langle-48,0\rangle}$ & & & & &$I$ & & & & & & & & & & & & & & & & & & & & & & & & \\\cline{2-33}
&8 & & &$T_{\langle-48,0\rangle}$ &$T_{\langle-42,0\rangle}$ & & & &$I$ &$T_{\langle6,0\rangle}$ & & & & & & & & & & & & & & & & & & & & & & \\\cline{2-33}
&9 & & &$T_{\langle-54,0\rangle}$ &$T_{\langle-48,0\rangle}$ & & & &$T_{\langle-6,0\rangle}$ &$I$ & & & & & & & & & & & & & & & & & & & & & & \\\cline{2-33}
&10 & & & & &$T_{\langle-48,0\rangle}$ & & & & &$I$ & & & & & & & & & & & & & & & & & & & & & \\\cline{2-33}
&11 & & & & & & & & & & &$I$ & & & & & & & & & & & & & & & & & & & & \\\cline{2-33}
&12 & & & & & & & & & & & &$I$ & & & & & &$T_{\langle60,0\rangle}$ & &$T_{\langle84,-7\rangle}$ & & & & & & & & & & & \\\cline{2-33}
&13 & & & & & & & & & & & & &$I$ & & & & & & & & & & & & & & & & & & \\\cline{2-33}
&14 & & & & & & & & & & & & & &$I$ & & & & & & & & & & & & & & & & & \\\cline{2-33}
&15 & & & & & & & & & & & & & & &$I$ & & & & & & & & & & & & & & & & \\\cline{2-33}
&16 & & & & & & & & & & & & & & & &$I$ & & &$T_{\langle24,-12\rangle}$ & & & & & & & & & & & & \\\cline{2-33}
&17 & & & & & & & & & & & & & & & & &$I$ & & & & & & & & & & & & & & \\\cline{2-33}
&18 & & & & & & & & & & & &$T_{\langle-60,0\rangle}$ & & & & & &$I$ & &$T_{\langle24,-7\rangle}$ & & & & & & & & & & & \\\cline{2-33}
&19 & & & & & & & & & & & & & & & &$T_{\langle-24,12\rangle}$ & & &$I$ & & & & & & & & & & & & \\\cline{2-33}
&20 & & & & & & & & & & & &$T_{\langle-84,7\rangle}$ & & & & & &$T_{\langle-24,7\rangle}$ & &$I$ & & & & & & & & & & & \\\cline{2-33}
&21 & & & & & & & & & & & & & & & & & & & & &$I$ & & & & & & & & & & \\\cline{2-33}
&22 & & & & & & & & & & & & & & & & & & & & & &$I$ & & & & & & & & & \\\cline{2-33}
&23 & & & & & & & & & & & & & & & & & & & & & & &$I$ & & & & & & & & \\\cline{2-33}
&24 & & & & & & & & & & & & & & & & & & & & & & & &$I$ & & & & & & & \\\cline{2-33}
&25 & & & & & & & & & & & & & & & & & & & & & & & & &$I$ & & & & & & \\\cline{2-33}
&26 & & & & & & & & & & & & & & & & & & & & & & & & & &$I$ &$T_{\langle6,0\rangle}$ & & & & \\\cline{2-33}
&27 & & & & & & & & & & & & & & & & & & & & & & & & & &$T_{\langle-6,0\rangle}$ &$I$ & & & & \\\cline{2-33}
&28 & & & & & & & & & & & & & & & & & & & & & & & & & & & &$I$ & & & \\\cline{2-33}
&A & & & & & & & & & & & & & & & & & & & & & & & & & & & & &$I$ & & \\\cline{2-33}
&B & & & & & & & & & & & & & & & & & & & & & & & & & & & & & &$I$ & \\\cline{2-33}
\multirow{-31}{*}{\rotatebox[origin=c]{90}{{\bfseries From pattern}}}&C & & & & & & & & & & & & & & & & & & & & & & & & & & & & & & &$I$ \\\cline{2-33}
\end{tabularx}
}
\end{sidewaystable}

\begin{sidewaystable}
\caption{Inter-pattern transformation graph for the set of patterns in the CPM representation in Fig.~\ref{RAVEL-HAYDN-GROUND-TRUTH-CPM-GT} and the transformation class $F_{\text{2STR}}$.}
\label{RAVEL-HAYDN-GROUND-TRUTH-CPM-IPTG}
\resizebox{\linewidth}{!}{
\begin{tabularx}{1.7\linewidth}{ll|lllllllllllllllllllllllllllllll}
&&\multicolumn{31}{c}{{\bfseries To pattern}}\\
&&1&2&3&4&5&6&7&8&9&10&11&12&13&14&15&16&17&18&19&20&21&22&23&24&25&26&27&28&A&B&C\\\cline{1-33}
&1 &$I$ & & & & &$T_{\langle96,0\rangle}$ & & & & & & & & & & & &$S_{-1}T_{\langle364,-12\rangle}$ & & & & & & & & & & & & & \\\cline{2-33}
&2 & &$I$ & & & & &$T_{\langle96,0\rangle}$ & & & & & & & & & & & & & & & & & & & & & & & & \\\cline{2-33}
&3 & & &$I$ &$T_{\langle12,0\rangle}$ & & & &$T_{\langle96,0\rangle}$ &$T_{\langle108,0\rangle}$ & & & & & & & & & & & & & & & & & & & & & & \\\cline{2-33}
&4 & & &$T_{\langle-12,0\rangle}$ &$I$ & & & &$T_{\langle84,0\rangle}$ &$T_{\langle96,0\rangle}$ & & & & & & & & & & & & & & & & & & & & & & \\\cline{2-33}
&5 & & & & &$I$ & & & & &$T_{\langle96,0\rangle}$ & & & & & & & & & & & & & & & & & & & & & \\\cline{2-33}
&6 &$T_{\langle-96,0\rangle}$ & & & & &$I$ & & & & & & & & & & & &$S_{-1}T_{\langle460,-12\rangle}$ & & & & & & & & & & & & & \\\cline{2-33}
&7 & &$T_{\langle-96,0\rangle}$ & & & & &$I$ & & & & & & & & & & & & & & & & & & & & & & & & \\\cline{2-33}
&8 & & &$T_{\langle-96,0\rangle}$ &$T_{\langle-84,0\rangle}$ & & & &$I$ &$T_{\langle12,0\rangle}$ & & & & & & & & & & & & & & & & & & & & & & \\\cline{2-33}
&9 & & &$T_{\langle-108,0\rangle}$ &$T_{\langle-96,0\rangle}$ & & & &$T_{\langle-12,0\rangle}$ &$I$ & & & & & & & & & & & & & & & & & & & & & & \\\cline{2-33}
&10 & & & & &$T_{\langle-96,0\rangle}$ & & & & &$I$ & & & & & & & & & & & & & & & & & & & & & \\\cline{2-33}
&11 & & & & & & & & & & &$I$ & & & & & & & & & & & & & & & & & & & & \\\cline{2-33}
&12 & & & & & & & & & & & &$I$ & & & & & & & &$T_{\langle168,-7\rangle}$ & & & & & & & & & & & \\\cline{2-33}
&13 & & & & & & & & & & & & &$I$ & & & & & & & & & & & & & & & & & & \\\cline{2-33}
&14 & & & & & & & & & & & & & &$I$ & & & & & & & & & & & & & & & & & \\\cline{2-33}
&15 & & & & & & & & & & & & & & &$I$ & & & & & & & & & & & & & & & & \\\cline{2-33}
&16 & & & & & & & & & & & & & & & &$I$ & & &$T_{\langle48,-12\rangle}$ & & & & & & & & & & &$S_{-1}T_{\langle720,-88\rangle}R_x$ & \\\cline{2-33}
&17 & & & & & & & & & & & & & & & & &$I$ & & & & & & & & & & & & & & \\\cline{2-33}
&18 &$S_{-1}T_{\langle364,12\rangle}$ & & & & &$S_{-1}T_{\langle460,12\rangle}$ & & & & & & & & & & & &$I$ & & & & & & & & & & & & & \\\cline{2-33}
&19 & & & & & & & & & & & & & & & &$T_{\langle-48,12\rangle}$ & & &$I$ & & & & & & & & & & &$S_{-1}T_{\langle768,-76\rangle}R_x$ & \\\cline{2-33}
&20 & & & & & & & & & & & &$T_{\langle-168,7\rangle}$ & & & & & & & &$I$ & & & & & & & & & & & \\\cline{2-33}
&21 & & & & & & & & & & & & & & & & & & & & &$I$ & & & & & & & & & & \\\cline{2-33}
&22 & & & & & & & & & & & & & & & & & & & & & &$I$ & & & & & & & & & \\\cline{2-33}
&23 & & & & & & & & & & & & & & & & & & & & & & &$I$ & & & & & & & & \\\cline{2-33}
&24 & & & & & & & & & & & & & & & & & & & & & & & &$I$ & & & & & & & \\\cline{2-33}
&25 & & & & & & & & & & & & & & & & & & & & & & & & &$I$ & & & & & & \\\cline{2-33}
&26 & & & & & & & & & & & & & & & & & & & & & & & & & &$I$ &$T_{\langle12,0\rangle}$ & & & & \\\cline{2-33}
&27 & & & & & & & & & & & & & & & & & & & & & & & & & &$T_{\langle-12,0\rangle}$ &$I$ & & & & \\\cline{2-33}
&28 & & & & & & & & & & & & & & & & & & & & & & & & & & & &$I$ & & & \\\cline{2-33}
&A & & & & & & & & & & & & & & & & & & & & & & & & & & & & &$I$ & & \\\cline{2-33}
&B & & & & & & & & & & & & & & & &$S_{-1}T_{\langle720,-88\rangle}R_x$ & & &$S_{-1}T_{\langle768,-76\rangle}R_x$ & & & & & & & & & & &$I$ & \\\cline{2-33}
\multirow{-31}{*}{\rotatebox[origin=c]{90}{{\bfseries From pattern}}}&C & & & & & & & & & & & & & & & & & & & & & & & & & & & & & & &$I$ \\\cline{2-33}
\end{tabularx}
}
\end{sidewaystable}

\begin{sidewaystable}
\caption{Inter-pattern transformation graph for the set of patterns in the MPO representation in Fig.~\ref{RAVEL-HAYDN-GROUND-TRUTH-MPO-GT} and the transformation class $F_{\text{2STR}}$.}
\label{RAVEL-HAYDN-GROUND-TRUTH-MPO-IPTG}
\resizebox{\linewidth}{!}{
\begin{tabularx}{1.715\linewidth}{ll|lllllllllllllllllllllllllllllll}
&&\multicolumn{31}{c}{{\bfseries To pattern}}\\
&&1&2&3&4&5&6&7&8&9&10&11&12&13&14&15&16&17&18&19&20&21&22&23&24&25&26&27&28&A&B&C\\\cline{1-33}
&1 &$I$ & & & & &$T_{\langle48,0\rangle}$ & & & & & & & & & & & & & & & & & & & & & & & &$T_{\langle190,-51\rangle}R_x$ & \\\cline{2-33}
&2 & &$I$ & & & & &$T_{\langle48,0\rangle}$ & & & & & & & & & & & & & & & & & & & & & & & & \\\cline{2-33}
&3 & & &$I$ &$T_{\langle6,0\rangle}$ & & & &$T_{\langle48,0\rangle}$ &$T_{\langle54,0\rangle}$ & & & & & & & & & & & & & & & & & & & & & & \\\cline{2-33}
&4 & & &$T_{\langle-6,0\rangle}$ &$I$ & & & &$T_{\langle42,0\rangle}$ &$T_{\langle48,0\rangle}$ & & & & & & & & & & & & & & & & & & & & & & \\\cline{2-33}
&5 & & & & &$I$ & & & & &$T_{\langle48,0\rangle}$ & & & & & & & & & & & & & & & & & & & & & \\\cline{2-33}
&6 &$T_{\langle-48,0\rangle}$ & & & & &$I$ & & & & & & & & & & & & & & & & & & & & & & & &$T_{\langle142,-51\rangle}R_x$ & \\\cline{2-33}
&7 & &$T_{\langle-48,0\rangle}$ & & & & &$I$ & & & & & & & & & & & & & & & & & & & & & & & & \\\cline{2-33}
&8 & & &$T_{\langle-48,0\rangle}$ &$T_{\langle-42,0\rangle}$ & & & &$I$ &$T_{\langle6,0\rangle}$ & & & & & & & & & & & & & & & & & & & & & & \\\cline{2-33}
&9 & & &$T_{\langle-54,0\rangle}$ &$T_{\langle-48,0\rangle}$ & & & &$T_{\langle-6,0\rangle}$ &$I$ & & & & & & & & & & & & & & & & & & & & & & \\\cline{2-33}
&10 & & & & &$T_{\langle-48,0\rangle}$ & & & & &$I$ & & & & & & & & & & & & & & & & & & & & & \\\cline{2-33}
&11 & & & & & & & & & & &$I$ & & & & & & & & & & & & & & & & & & & & \\\cline{2-33}
&12 & & & & & & & & & & & &$I$ & &$T_{\langle36,-51\rangle}R_x$ & &$T_{\langle48,7\rangle}$ & &$T_{\langle60,0\rangle}$ &$T_{\langle72,0\rangle}$ &$T_{\langle84,-4\rangle}$ & & & & & & & & & & & \\\cline{2-33}
&13 & & & & & & & & & & & & &$I$ & & & & & & & & & & & & & & & & & & \\\cline{2-33}
&14 & & & & & & & & & & & &$T_{\langle-36,-51\rangle}R_x$ & &$I$ & &$T_{\langle12,-58\rangle}R_x$ & &$T_{\langle24,-51\rangle}R_x$ &$T_{\langle36,-51\rangle}R_x$ &$T_{\langle48,-47\rangle}R_x$ & & & & & & & & & & & \\\cline{2-33}
&15 & & & & & & & & & & & & & & &$I$ & & & & & & & & & & & & & & & & \\\cline{2-33}
&16 & & & & & & & & & & & &$T_{\langle-48,-7\rangle}$ & &$T_{\langle-12,-58\rangle}R_x$ & &$I$ & &$T_{\langle12,-7\rangle}$ &$T_{\langle24,-7\rangle}$ &$T_{\langle36,-11\rangle}$ & & & & & & & & & & & \\\cline{2-33}
&17 & & & & & & & & & & & & & & & & &$I$ & & & & & & & & & & & & & & \\\cline{2-33}
&18 & & & & & & & & & & & &$T_{\langle-60,0\rangle}$ & &$T_{\langle-24,-51\rangle}R_x$ & &$T_{\langle-12,7\rangle}$ & &$I$ &$T_{\langle12,0\rangle}$ &$T_{\langle24,-4\rangle}$ & & & & & & & & & & & \\\cline{2-33}
&19 & & & & & & & & & & & &$T_{\langle-72,0\rangle}$ & &$T_{\langle-36,-51\rangle}R_x$ & &$T_{\langle-24,7\rangle}$ & &$T_{\langle-12,0\rangle}$ &$I$ &$T_{\langle12,-4\rangle}$ & & & & & & & & & & & \\\cline{2-33}
&20 & & & & & & & & & & & &$T_{\langle-84,4\rangle}$ & &$T_{\langle-48,-47\rangle}R_x$ & &$T_{\langle-36,11\rangle}$ & &$T_{\langle-24,4\rangle}$ &$T_{\langle-12,4\rangle}$ &$I$ & & & & & & & & & & & \\\cline{2-33}
&21 & & & & & & & & & & & & & & & & & & & & &$I$ & & & & & & & & & & \\\cline{2-33}
&22 & & & & & & & & & & & & & & & & & & & & & &$I$ & & & & & & & & & \\\cline{2-33}
&23 & & & & & & & & & & & & & & & & & & & & & & &$I$ & & & & & & & & \\\cline{2-33}
&24 & & & & & & & & & & & & & & & & & & & & & & & &$I$ & & & & & & & \\\cline{2-33}
&25 & & & & & & & & & & & & & & & & & & & & & & & & &$I$ & & & & & & \\\cline{2-33}
&26 & & & & & & & & & & & & & & & & & & & & & & & & & &$I$ &$T_{\langle6,0\rangle}$ & & & & \\\cline{2-33}
&27 & & & & & & & & & & & & & & & & & & & & & & & & & &$T_{\langle-6,0\rangle}$ &$I$ & & & & \\\cline{2-33}
&28 & & & & & & & & & & & & & & & & & & & & & & & & & & & &$I$ & & & \\\cline{2-33}
&A & & & & & & & & & & & & & & & & & & & & & & & & & & & & &$I$ & & \\\cline{2-33}
&B &$T_{\langle-190,-51\rangle}R_x$ & & & & &$T_{\langle-142,-51\rangle}R_x$ & & & & & & & & & & & & & & & & & & & & & & & &$I$ & \\\cline{2-33}
\multirow{-31}{*}{\rotatebox[origin=c]{90}{{\bfseries From pattern}}}&C & & & & & & & & & & & & & & & & & & & & & & & & & & & & & & &$I$ \\\cline{2-33}
\end{tabularx}
}
\end{sidewaystable}

\begin{sidewaystable}
\caption{Inter-pattern transformation graph for the set of patterns in the MPM representation in Fig.~\ref{RAVEL-HAYDN-GROUND-TRUTH-MPM-GT} and the transformation class $F_{\text{2STR}}$.}
\label{RAVEL-HAYDN-GROUND-TRUTH-MPM-IPTG}
\resizebox{\linewidth}{!}{
\begin{tabularx}{1.86\linewidth}{ll|lllllllllllllllllllllllllllllll}
&&\multicolumn{31}{c}{{\bfseries To pattern}}\\
&&1&2&3&4&5&6&7&8&9&10&11&12&13&14&15&16&17&18&19&20&21&22&23&24&25&26&27&28&A&B&C\\\cline{1-33}
&1 &$I$ & & & & &$T_{\langle96,0\rangle}$ & & & & & & & &$S_{-1}T_{\langle316,-58\rangle}R_x$ & &$S_{-1}T_{\langle340,0\rangle}$ & &$S_{-1}T_{\langle364,-7\rangle}$ &$S_{-1}T_{\langle388,-7\rangle}$ & & & & & & & & & & &$T_{\langle380,-51\rangle}R_x$ & \\\cline{2-33}
&2 & &$I$ & & & & &$T_{\langle96,0\rangle}$ & & & & & & & & & & & & & & & & & & & & & & & & \\\cline{2-33}
&3 & & &$I$ &$T_{\langle12,0\rangle}$ & & & &$T_{\langle96,0\rangle}$ &$T_{\langle108,0\rangle}$ & & & & & & & & & & & & & & & & & & & & & & \\\cline{2-33}
&4 & & &$T_{\langle-12,0\rangle}$ &$I$ & & & &$T_{\langle84,0\rangle}$ &$T_{\langle96,0\rangle}$ & & & & & & & & & & & & & & & & & & & & & & \\\cline{2-33}
&5 & & & & &$I$ & & & & &$T_{\langle96,0\rangle}$ & & & & & & & & & & & & & & & & & & & & & \\\cline{2-33}
&6 &$T_{\langle-96,0\rangle}$ & & & & &$I$ & & & & & & & &$S_{-1}T_{\langle412,-58\rangle}R_x$ & &$S_{-1}T_{\langle436,0\rangle}$ & &$S_{-1}T_{\langle460,-7\rangle}$ &$S_{-1}T_{\langle484,-7\rangle}$ & & & & & & & & & & &$T_{\langle284,-51\rangle}R_x$ & \\\cline{2-33}
&7 & &$T_{\langle-96,0\rangle}$ & & & & &$I$ & & & & & & & & & & & & & & & & & & & & & & & & \\\cline{2-33}
&8 & & &$T_{\langle-96,0\rangle}$ &$T_{\langle-84,0\rangle}$ & & & &$I$ &$T_{\langle12,0\rangle}$ & & & & & & & & & & & & & & & & & & & & & & \\\cline{2-33}
&9 & & &$T_{\langle-108,0\rangle}$ &$T_{\langle-96,0\rangle}$ & & & &$T_{\langle-12,0\rangle}$ &$I$ & & & & & & & & & & & & & & & & & & & & & & \\\cline{2-33}
&10 & & & & &$T_{\langle-96,0\rangle}$ & & & & &$I$ & & & & & & & & & & & & & & & & & & & & & \\\cline{2-33}
&11 & & & & & & & & & & &$I$ & & & & & & & & & & & & & & & & & & & & \\\cline{2-33}
&12 & & & & & & & & & & & &$I$ & & & & & & & &$T_{\langle168,-4\rangle}$ & & & & & & & & & & & \\\cline{2-33}
&13 & & & & & & & & & & & & &$I$ & & & & & & & & & & & & & & & & & & \\\cline{2-33}
&14 &$S_{-1}T_{\langle316,-58\rangle}R_x$ & & & & &$S_{-1}T_{\langle412,-58\rangle}R_x$ & & & & & & & &$I$ & &$T_{\langle24,-58\rangle}R_x$ & &$T_{\langle48,-51\rangle}R_x$ &$T_{\langle72,-51\rangle}R_x$ & & & & & & & & & & &$S_{-1}T_{\langle696,-7\rangle}$ & \\\cline{2-33}
&15 & & & & & & & & & & & & & & &$I$ & & & & & & & & & & & & & & & & \\\cline{2-33}
&16 &$S_{-1}T_{\langle340,0\rangle}$ & & & & &$S_{-1}T_{\langle436,0\rangle}$ & & & & & & & &$T_{\langle-24,-58\rangle}R_x$ & &$I$ & &$T_{\langle24,-7\rangle}$ &$T_{\langle48,-7\rangle}$ & & & & & & & & & & &$S_{-1}T_{\langle720,-51\rangle}R_x$ & \\\cline{2-33}
&17 & & & & & & & & & & & & & & & & &$I$ & & & & & & & & & & & & & & \\\cline{2-33}
&18 &$S_{-1}T_{\langle364,7\rangle}$ & & & & &$S_{-1}T_{\langle460,7\rangle}$ & & & & & & & &$T_{\langle-48,-51\rangle}R_x$ & &$T_{\langle-24,7\rangle}$ & &$I$ &$T_{\langle24,0\rangle}$ & & & & & & & & & & &$S_{-1}T_{\langle744,-44\rangle}R_x$ & \\\cline{2-33}
&19 &$S_{-1}T_{\langle388,7\rangle}$ & & & & &$S_{-1}T_{\langle484,7\rangle}$ & & & & & & & &$T_{\langle-72,-51\rangle}R_x$ & &$T_{\langle-48,7\rangle}$ & &$T_{\langle-24,0\rangle}$ &$I$ & & & & & & & & & & &$S_{-1}T_{\langle768,-44\rangle}R_x$ & \\\cline{2-33}
&20 & & & & & & & & & & & &$T_{\langle-168,4\rangle}$ & & & & & & & &$I$ & & & & & & & & & & & \\\cline{2-33}
&21 & & & & & & & & & & & & & & & & & & & & &$I$ & & & & & & & & & & \\\cline{2-33}
&22 & & & & & & & & & & & & & & & & & & & & & &$I$ & & & & & & & & & \\\cline{2-33}
&23 & & & & & & & & & & & & & & & & & & & & & & &$I$ & & & & & & & & \\\cline{2-33}
&24 & & & & & & & & & & & & & & & & & & & & & & & &$I$ & & & & & & & \\\cline{2-33}
&25 & & & & & & & & & & & & & & & & & & & & & & & & &$I$ & & & & & & \\\cline{2-33}
&26 & & & & & & & & & & & & & & & & & & & & & & & & & &$I$ &$T_{\langle12,0\rangle}$ & & & & \\\cline{2-33}
&27 & & & & & & & & & & & & & & & & & & & & & & & & & &$T_{\langle-12,0\rangle}$ &$I$ & & & & \\\cline{2-33}
&28 & & & & & & & & & & & & & & & & & & & & & & & & & & & &$I$ & & & \\\cline{2-33}
&A & & & & & & & & & & & & & & & & & & & & & & & & & & & & &$I$ & & \\\cline{2-33}
&B &$T_{\langle-380,-51\rangle}R_x$ & & & & &$T_{\langle-284,-51\rangle}R_x$ & & & & & & & &$S_{-1}T_{\langle696,7\rangle}$ & &$S_{-1}T_{\langle720,-51\rangle}R_x$ & &$S_{-1}T_{\langle744,-44\rangle}R_x$ &$S_{-1}T_{\langle768,-44\rangle}R_x$ & & & & & & & & & & &$I$ & \\\cline{2-33}
\multirow{-31}{*}{\rotatebox[origin=c]{90}{{\bfseries From pattern}}}&C & & & & & & & & & & & & & & & & & & & & & & & & & & & & & & &$I$ \\\cline{2-33}
\end{tabularx}
}
\end{sidewaystable}

\subsection{Inter-pattern transformation graphs}

We have seen that many of the occurrences of the HAYDN theme in Ravel's {\em Menuet\/} are related to each other by combinations of the traditional contrapuntal transformations of transposition, inversion and retrograde and that these musical transformations correspond, respectively, to translation, reflection in the $x$-axis and reflection in the $y$-axis in one or more of the four types of geometric pitch--time representation introduced above---that is, the CPO, CPM, MPO and MPM representations. 

Given a set of patterns, $Q$, in one of these types of representation, along with a class of transformations, $\tc$, we can use a graph, represented as a table, to tell us which transformations in $\tc$ map which patterns in $Q$ completely and exactly onto other patterns in $Q$. Table~\ref{RAVEL-HAYDN-GROUND-TRUTH-CPO-IPTG} shows such a graph for the set of patterns in the CPO representation in Fig.~\ref{RAVEL-HAYDN-GROUND-TRUTH-CPO-GT} and the class of transformations that consist of a scaling parallel to the $x$-axis by a scale factor $f$, which may be any real value (denoted by $S_f$), followed by a translation by a vector $v$ (denoted by $T_v$), optionally followed by a reflection in the $x$-axis (denoted by $R_x$). The identity transformation is denoted by $I$. From this point on, we will refer to this class of transformations as $F_{\text{2STR}}$.
We call such a graph an {\em inter-pattern transformation graph\/} (IPTG) and the corresponding tabular representation is essentially an adjacency matrix \citep[pp.~589--592]{CormenEtAl2009}, where an edge from pattern $P_1$ to pattern $P_2$ is labelled with the transformation(s) in the specified transformation class that map $P_1$ onto $P_2$.

As can be seen in Table~\ref{RAVEL-HAYDN-GROUND-TRUTH-CPO-IPTG}, all the cases in Fig.~\ref{RAVEL-HAYDN-GROUND-TRUTH-CPO-GT} where different occurrences of the HAYDN theme are related by transformations in $F_{\text{2STR}}$ are cases where the patterns are simply translationally equivalent. That is, there is no pattern amongst those indicated in Fig.~\ref{RAVEL-HAYDN-GROUND-TRUTH-CPO-GT} that is mapped exactly and completely onto another pattern in this representation by a transformation in $F_{\text{2STR}}$ that involves scaling parallel to or reflection in the $x$-axis. By reading along the rows (or down the columns) in Table~\ref{RAVEL-HAYDN-GROUND-TRUTH-CPO-IPTG} we can easily confirm that the following sets of two or more occurrences in the CPO representation of Ravel's {\em Menuet\/} form sets of translationally equivalent patterns: $\lbrace1,6\rbrace$, $\lbrace2,7\rbrace$, $\lbrace3,4,8,9\rbrace$, $\lbrace5,10\rbrace$, $\lbrace12,18,20\rbrace$, $\lbrace16,19\rbrace$ and $\lbrace26,27\rbrace$. 

Table~\ref{RAVEL-HAYDN-GROUND-TRUTH-CPM-IPTG} shows the inter-pattern transformation graph for the occurrences of the HAYDN theme indicated in the CPM representation in Fig.~\ref{RAVEL-HAYDN-GROUND-TRUTH-CPM-GT} and the transformation class $F_{\text{2STR}}$. Again, because every transformation in Table~\ref{RAVEL-HAYDN-GROUND-TRUTH-CPM-IPTG} is invertible and the inverse of every invertible transformation in $F_{\text{2STR}}$ is also in $F_{\text{2STR}}$, we can simply read down the columns or across the rows to determine the sets of patterns in Fig.~\ref{RAVEL-HAYDN-GROUND-TRUTH-CPM-GT} that are transformationally equivalent with respect to the transformation class $F_{\text{2STR}}$. For example,  rows 1, 6 and 18 in the table collectively tell us that occurrences 1, 6 and 18 form a set of transformationally equivalent patterns in which each pattern can be mapped exactly and completely onto the other patterns in the set by transformations in $F_{\text{2STR}}$. Note that, as discussed above, in the CPM representation, there exist pairs of occurrences that are related by transformations that involve reflection in the $x$- and $y$-axes (corresponding to inversion and retrograde, respectively). For example, pattern 16 is mapped onto pattern B by the transformation denoted in Table~\ref{RAVEL-HAYDN-GROUND-TRUTH-CPM-IPTG} by $S_{-1}T_{\langle720,-88\rangle}R_x$, which consists of a reflection in the $y$-axis (equivalent to a scaling parallel to the $x$-axis by a scale factor of $-1$, denoted by $S_{-1}$), followed by a translation by the vector $\langle720,-88\rangle$ (denoted by $T_{\langle720,-88\rangle}$), followed by a reflection in the $x$-axis (denoted by $R_x$).  From Table~\ref{RAVEL-HAYDN-GROUND-TRUTH-CPM-IPTG} we can determine that the following sets of HAYDN occurrences in the CPM representation form sets of two or more patterns that are transformationally equivalent with respect to $F_{\text{2STR}}$: $\lbrace1,6,18\rbrace$, $\lbrace2,7\rbrace$, $\lbrace3,4,8,9\rbrace$, $\lbrace5,10\rbrace$, $\lbrace12,20\rbrace$, $\lbrace16,19,\text{B}\rbrace$ and $\lbrace26,27\rbrace$.

Table~\ref{RAVEL-HAYDN-GROUND-TRUTH-MPO-IPTG} shows the inter-pattern transformation graph for the occurrences of the HAYDN theme indicated in the MPO representation in Fig.~\ref{RAVEL-HAYDN-GROUND-TRUTH-MPO-GT} and the transformation class $F_{\text{2STR}}$. This table tells us that the following sets of HAYDN occurrences in the MPO representation form sets of two or more patterns that are transformationally equivalent with respect to $F_{\text{2STR}}$: $\lbrace1,6,\text{B}\rbrace$, $\lbrace2,7\rbrace$, $\lbrace3,4,8,9\rbrace$, $\lbrace5,10\rbrace$, $\lbrace12,14,16,18,19,20\rbrace$ and $\lbrace26,27\rbrace$.

Finally, Table~\ref{RAVEL-HAYDN-GROUND-TRUTH-MPM-IPTG} shows the inter-pattern transformation graph for the occurrences of the HAYDN theme indicated in the MPM representation in Fig.~\ref{RAVEL-HAYDN-GROUND-TRUTH-MPM-GT} and the transformation class $F_{\text{2STR}}$. This table tells us that the following sets of HAYDN occurrences in the MPM representation form sets of two or more patterns that are transformationally equivalent with respect to $F_{\text{2STR}}$: $\lbrace1,6,14,16,18,19,\text{B}\rbrace$, $\lbrace2,7\rbrace$, $\lbrace3,4,8,9\rbrace$, $\lbrace5,10\rbrace$, $\lbrace12,20\rbrace$ and $\lbrace26,27\rbrace$.

\subsection{Chroma--time and morph--time representations}

Examination of the inter-pattern transformation graphs in Tables~\ref{RAVEL-HAYDN-GROUND-TRUTH-CPO-IPTG}--\ref{RAVEL-HAYDN-GROUND-TRUTH-MPM-IPTG} reveals that there are twelve occurrences of the HAYDN theme identified in Figs.~\ref{figure-soucy-haydn-occurrences} and \ref{further-occurrences-of-HAYDN-theme} that cannot be mapped by any transformation in the class, $F_{\text{2STR}}$, in any of the four representations shown in Figs.~\ref{RAVEL-HAYDN-GROUND-TRUTH-CPO-GT}--\ref{RAVEL-HAYDN-GROUND-TRUTH-MPM-GT} onto any of the other occurrences of the theme. These are occurrences 11, 13, 15, 17, 21--25, 28, A and C. Moreover, none of the other occurrences can be mapped onto any of these twelve occurrences by any transformation in $F_{\text{2STR}}$ in any of the four types of representation considered thus far. In the CPO representation, occurrences 14 and B are also unrelated to any of the other occurrences with respect to $F_{\text{2STR}}$; and in the CPM representation, occurrence 14 is unrelated to the other occurrences with respect to $F_{\text{2STR}}$.

If $p_{\text c}$ is the chromatic pitch of a note, then, following \citet[p.~127]{Meredith2006}, we define the {\em chroma}, $c$, of the note to be given by $c=p_{\text c}\bmod 12$.\footnote{Note that we define that $a \bmod b = a - b\left\lfloor\frac a b\right\rfloor$.} Similarly, if $p_{\text m}$ is the morphetic pitch of a note, then, again following \citet[p.~127]{Meredith2006}, we define the {\em morph}, $m$, of the note to be given by $m=p_{\text m}\bmod 7$. 

\begin{sidewaysfigure}
\resizebox{\linewidth}{!}{\includegraphics{"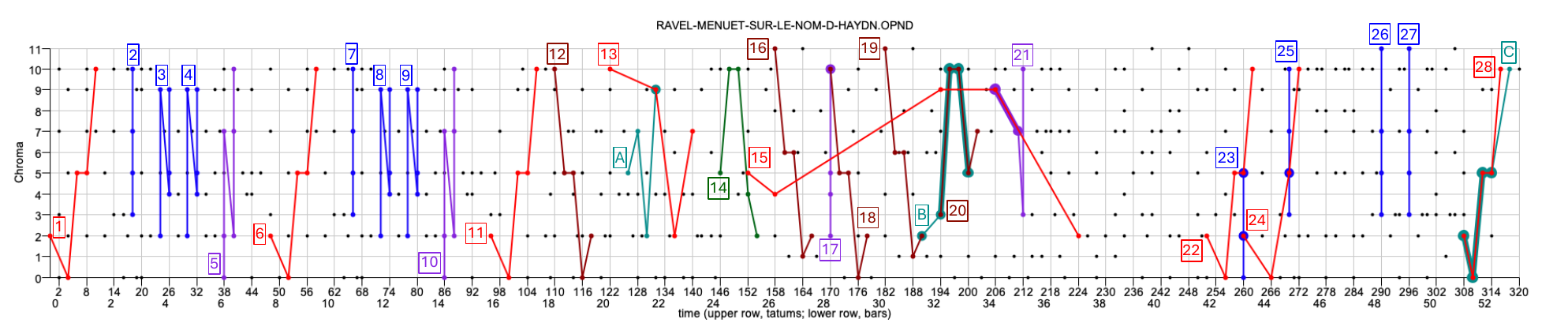"}}
\caption{A {\em chroma--onset\/} or CO representation of Ravel's {\em Menuet sur le nom d'Haydn}. Each point represents one or more tones in the score. The $x$-coordinate of a point gives the onset time in tatums (in this case, quavers) of the tones represented by the point, and the $y$--coordinate gives the chroma of the tones represented by the point. The occurrences of the HAYDN theme identified in Figs.~\ref{figure-soucy-haydn-occurrences} and \ref{further-occurrences-of-HAYDN-theme} are shown, labelled with the same numbers and letters and coloured using the same colours as in these previous figures. When a point occurs in two patterns, it is drawn twice with different diameters and colours so that its presence in both patterns is visible.}
	\label{RAVEL-HAYDN-GROUND-TRUTH-CO-GT}
\end{sidewaysfigure}

\begin{sidewaystable}
\caption{The inter-pattern transformation graph for the set of HAYDN occurrences in the CO representation shown in Fig.~\ref{RAVEL-HAYDN-GROUND-TRUTH-CO-GT} and the transformation class $F_{\mathrm{2STRmod12}}$.}
\label{RAVEL-HAYDN-GROUND-TRUTH-CO-IPTG}
\resizebox{\linewidth}{!}{
\begin{tabularx}{1.73\linewidth}{ll|lllllllllllllllllllllllllllllll}
&&\multicolumn{31}{c}{{\bfseries To pattern}}\\
&&1&2&3&4&5&6&7&8&9&10&11&12&13&14&15&16&17&18&19&20&21&22&23&24&25&26&27&28&A&B&C\\\cline{1-33}
&1 &$I$ &$S_{0}T_{\langle18,5\rangle}$ & & & &$T_{\langle48,0\rangle}$ &$S_{0}T_{\langle66,5\rangle}$ & & & &$T_{\langle96,0\rangle}$ & & & & & & & & & & &$T_{\langle252,0\rangle}$ & & &$S_{0}T_{\langle270,5\rangle}$ & & & & & & \\\cline{2-33}
&2 & &$I$ & & & & &$T_{\langle48,0\rangle}$ & & & & & & & & & & & & & & & & & &$T_{\langle252,0\rangle}$ & & & & & & \\\cline{2-33}
&3 & & &$I$ &$T_{\langle6,0\rangle}$ & & & &$T_{\langle48,0\rangle}$ &$T_{\langle54,0\rangle}$ & & & & & & & & & & & & & & & & & & & & & & \\\cline{2-33}
&4 & & &$T_{\langle-6,0\rangle}$ &$I$ & & & &$T_{\langle42,0\rangle}$ &$T_{\langle48,0\rangle}$ & & & & & & & & & & & & & & & & & & & & & & \\\cline{2-33}
&5 & &$S_{0}T_{\langle18,7\rangle}R_x$ & & &$I$ & &$S_{0}T_{\langle66,7\rangle}R_x$ & & &$T_{\langle48,0\rangle}$ & & & & & & & & & & & & & & &$S_{0}T_{\langle270,7\rangle}R_x$ & & & & & & \\\cline{2-33}
&6 &$T_{\langle-48,0\rangle}$ &$S_{0}T_{\langle18,5\rangle}$ & & & &$I$ &$S_{0}T_{\langle66,5\rangle}$ & & & &$T_{\langle48,0\rangle}$ & & & & & & & & & & &$T_{\langle204,0\rangle}$ & & &$S_{0}T_{\langle270,5\rangle}$ & & & & & & \\\cline{2-33}
&7 & &$T_{\langle-48,0\rangle}$ & & & & &$I$ & & & & & & & & & & & & & & & & & &$T_{\langle204,0\rangle}$ & & & & & & \\\cline{2-33}
&8 & & &$T_{\langle-48,0\rangle}$ &$T_{\langle-42,0\rangle}$ & & & &$I$ &$T_{\langle6,0\rangle}$ & & & & & & & & & & & & & & & & & & & & & & \\\cline{2-33}
&9 & & &$T_{\langle-54,0\rangle}$ &$T_{\langle-48,0\rangle}$ & & & &$T_{\langle-6,0\rangle}$ &$I$ & & & & & & & & & & & & & & & & & & & & & & \\\cline{2-33}
&10 & &$S_{0}T_{\langle18,7\rangle}R_x$ & & &$T_{\langle-48,0\rangle}$ & &$S_{0}T_{\langle66,7\rangle}R_x$ & & &$I$ & & & & & & & & & & & & & & &$S_{0}T_{\langle270,7\rangle}R_x$ & & & & & & \\\cline{2-33}
&11 &$T_{\langle-96,0\rangle}$ &$S_{0}T_{\langle18,5\rangle}$ & & & &$T_{\langle-48,0\rangle}$ &$S_{0}T_{\langle66,5\rangle}$ & & & &$I$ & & & & & & & & & & &$T_{\langle156,0\rangle}$ & & &$S_{0}T_{\langle270,5\rangle}$ & & & & & & \\\cline{2-33}
&12 & &$S_{0}T_{\langle18,5\rangle}$ & & & & &$S_{0}T_{\langle66,5\rangle}$ & & & & &$I$ & & & & & &$T_{\langle60,0\rangle}$ & &$T_{\langle84,5\rangle}$ & & & & &$S_{0}T_{\langle270,5\rangle}$ & & &$S_{-1}T_{\langle426,0\rangle}$ & & & \\\cline{2-33}
&13 & & & & & & & & & & & & &$I$ & & & & & & & & & & & & & & & & & & \\\cline{2-33}
&14 & & & & & & & & & & & & & &$I$ & & &$S_{0}T_{\langle170,0\rangle}$ & & & & & & & & & & & & & & \\\cline{2-33}
&15 & & & & & & & & & & & & & & &$I$ & & & & & & & & & & & & & & & & \\\cline{2-33}
&16 & & & & & & & & & & & & & & & &$I$ & & &$T_{\langle24,0\rangle}$ & & & & & & & & & & & & \\\cline{2-33}
&17 & & & & & & & & & & & & & & & & &$I$ & & & & & & & & & & & & & & \\\cline{2-33}
&18 & &$S_{0}T_{\langle18,5\rangle}$ & & & & &$S_{0}T_{\langle66,5\rangle}$ & & & & &$T_{\langle-60,0\rangle}$ & & & & & &$I$ & &$T_{\langle24,5\rangle}$ & & & & &$S_{0}T_{\langle270,5\rangle}$ & & &$S_{-1}T_{\langle486,0\rangle}$ & & & \\\cline{2-33}
&19 & & & & & & & & & & & & & & & &$T_{\langle-24,0\rangle}$ & & &$I$ & & & & & & & & & & & & \\\cline{2-33}
&20 & &$S_{0}T_{\langle18,0\rangle}$ & & & & &$S_{0}T_{\langle66,0\rangle}$ & & & & &$T_{\langle-84,7\rangle}$ & & & & & &$T_{\langle-24,7\rangle}$ & &$I$ & & & & &$S_{0}T_{\langle270,0\rangle}$ & & &$S_{-1}T_{\langle510,7\rangle}$ & & & \\\cline{2-33}
&21 & & & & & & & & & & & & & & & & &$S_{0}T_{\langle170,7\rangle}$ & & & &$I$ & & & & & & & & & & \\\cline{2-33}
&22 &$T_{\langle-252,0\rangle}$ &$S_{0}T_{\langle18,5\rangle}$ & & & &$T_{\langle-204,0\rangle}$ &$S_{0}T_{\langle66,5\rangle}$ & & & &$T_{\langle-156,0\rangle}$ & & & & & & & & & & &$I$ & & &$S_{0}T_{\langle270,5\rangle}$ & & & & & & \\\cline{2-33}
&23 & & & & & & & & & & & & & & & & & & & & & & &$I$ & & & & & & & & \\\cline{2-33}
&24 & &$S_{0}T_{\langle18,5\rangle}$ & & & & &$S_{0}T_{\langle66,5\rangle}$ & & & & & & & & & & & & & & & & &$I$ &$S_{0}T_{\langle270,5\rangle}$ & & & & & & \\\cline{2-33}
&25 & &$T_{\langle-252,0\rangle}$ & & & & &$T_{\langle-204,0\rangle}$ & & & & & & & & & & & & & & & & & &$I$ & & & & & & \\\cline{2-33}
&26 & & & & & & & & & & & & & & & & & & & & & & & & & &$I$ &$T_{\langle6,0\rangle}$ & & & & \\\cline{2-33}
&27 & & & & & & & & & & & & & & & & & & & & & & & & & &$T_{\langle-6,0\rangle}$ &$I$ & & & & \\\cline{2-33}
&28 & &$S_{0}T_{\langle18,5\rangle}$ & & & & &$S_{0}T_{\langle66,5\rangle}$ & & & & &$S_{-1}T_{\langle426,0\rangle}$ & & & & & &$S_{-1}T_{\langle486,0\rangle}$ & &$S_{-1}T_{\langle510,5\rangle}$ & & & & &$S_{0}T_{\langle270,5\rangle}$ & & &$I$ & & & \\\cline{2-33}
&A & &$S_{0}T_{\langle18,0\rangle}R_x$ & & & & &$S_{0}T_{\langle66,0\rangle}R_x$ & & & & & & & & & & & & & & & & & &$S_{0}T_{\langle270,0\rangle}R_x$ & & & &$I$ & & \\\cline{2-33}
&B & & & & & & & & & & & & & & & & & & & & & & & & & & & & & &$I$ & \\\cline{2-33}
\multirow{-31}{*}{\rotatebox[origin=c]{90}{{\bfseries From pattern}}}&C & &$S_{0}T_{\langle18,5\rangle}$ & & & & &$S_{0}T_{\langle66,5\rangle}$ & & & & & & & & & & & & & & & & & &$S_{0}T_{\langle270,5\rangle}$ & & & & & &$I$ \\\cline{2-33}
\end{tabularx}
}
\end{sidewaystable}

Figure~\ref{RAVEL-HAYDN-GROUND-TRUTH-CO-GT} shows a 2-dimensional point set representing Ravel's {\em Menuet}, in which each point, $\langle t,c\rangle$, represents at least one tone in the score (in the sense of footnote~\ref{footnote-note-and-tone}), and where $t$ and $c$ are the onset time in tatums and the chroma, respectively, of the tones represented by that point. The occurrences of the HAYDN theme identified in Figs.~\ref{figure-soucy-haydn-occurrences} and \ref{further-occurrences-of-HAYDN-theme} are shown, labelled with the same numbers and letters and coloured using the same colours as in these previous figures. We refer to this type of representation as a {\em chroma--onset\/} (or CO) representation. Table~\ref{RAVEL-HAYDN-GROUND-TRUTH-CO-IPTG} shows the inter-pattern transformation graph for the patterns indicated in this CO representation and the class of 2-dimensional transformations consisting of a scaling parallel to the $x$-axis, followed by a translation by a vector where the $y$-component of the vector is added $\bmod 12$ to the $y$-component of each point, optionally followed by a reflection in the $x$-axis (also $\bmod 12$). We denote this transformation class by $F_{\text{2STRmod12}}$.

From Table~\ref{RAVEL-HAYDN-GROUND-TRUTH-CO-IPTG} we can see that every HAYDN occurrence identified in Figs.~\ref{figure-soucy-haydn-occurrences} and \ref{further-occurrences-of-HAYDN-theme}, except occurrences 13, 15, 23 and B is related to at least one other occurrence by a transformation in $F_{\text{2STRmod12}}$ in the CO representation of the {\em Menuet}. In particular, note that 13 occurrences are mapped onto the ``chordal'' occurrences 2, 7 and 25 by transformations that involve an $x$-axis scaling by a scale factor of 0. These transformations are not invertible since there is no unique inverse of an $x$-axis scaling with scale factor 0. Consequently, the set of occurrences that are mapped onto occurrences 2, 7 and 25 do not form a set of transformationally equivalent patterns with respect to $F_{\text{2STRmod12}}$, since some of the patterns within this set cannot be mapped onto other patterns in the set---for example, occurrence 6 can be mapped onto occurrence 2, but occurrence 2 cannot be mapped onto occurrence 6. The CO representation is particularly useful for discovering the relationships between occurrences in which pitches are displaced to different octaves, such as occurrences 1, 11 and 22---note that, of the types of representation discussed so far, only the CO representation allows for the geometric relationships between these three patterns to be found (examine the row for occurrence 1 in Tables~\ref{RAVEL-HAYDN-GROUND-TRUTH-CPO-IPTG}--\ref{RAVEL-HAYDN-GROUND-TRUTH-MPM-IPTG}).

\begin{sidewaysfigure}
\resizebox{\linewidth}{!}{\includegraphics{"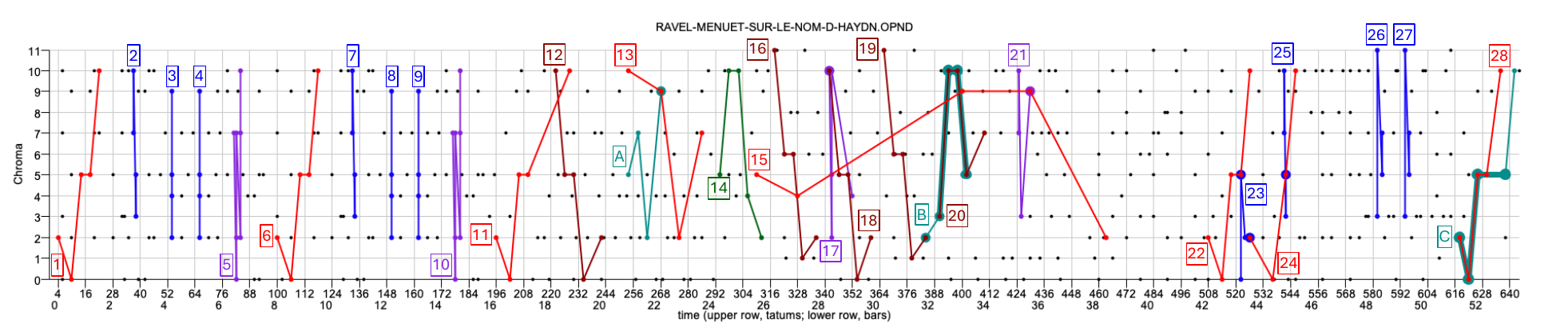"}}
\caption{A {\em chroma--midtime\/} or CM representation of Ravel's {\em Menuet sur le nom d'Haydn}. Each point represents one or more tones in the score. The $x$-coordinate of a point gives the midtime in tatums (in this case, semiquavers) of the tones represented by the point, and the $y$--coordinate gives the chroma of the tones represented by the point. The occurrences of the HAYDN theme identified in Figs.~\ref{figure-soucy-haydn-occurrences} and \ref{further-occurrences-of-HAYDN-theme} are shown, labelled with the same numbers and letters and coloured using the same colours as in these previous figures. When a point occurs in two patterns, it is drawn twice with different diameters and colours so that its presence in both patterns is visible.}
	\label{RAVEL-HAYDN-GROUND-TRUTH-CM-GT}
\end{sidewaysfigure}

\begin{sidewaystable}
\caption{The inter-pattern transformation graph for the patterns indicated in  the CM representation in Fig.~\ref{RAVEL-HAYDN-GROUND-TRUTH-CM-GT} and the transformation class $F_{\text{2STRmod12}}$.}
\label{RAVEL-HAYDN-GROUND-TRUTH-CM-IPTG}
\resizebox{\linewidth}{!}{
\begin{tabularx}{1.95\linewidth}{ll|lllllllllllllllllllllllllllllll}
&&\multicolumn{31}{c}{{\bfseries To pattern}}\\
&&1&2&3&4&5&6&7&8&9&10&11&12&13&14&15&16&17&18&19&20&21&22&23&24&25&26&27&28&A&B&C\\\cline{1-33}
&1 &$I$ & & & & &$T_{\langle96,0\rangle}$ & & & & & & & & & & & &$S_{-1}T_{\langle364,0\rangle}$ & & & &$T_{\langle504,0\rangle}$ & & & & & & & & & \\\cline{2-33}
&2 & &$I$ & & &$S_{-1}T_{\langle120,7\rangle}R_x$ & &$T_{\langle96,0\rangle}$ & & &$S_{-1}T_{\langle216,7\rangle}R_x$ & & & & & & & & & & & & & & &$T_{\langle504,0\rangle}$ & & & & & & \\\cline{2-33}
&3 & & &$I$ &$T_{\langle12,0\rangle}$ & & & &$T_{\langle96,0\rangle}$ &$T_{\langle108,0\rangle}$ & & & & & & & & & & & & & & & & & & & & & & \\\cline{2-33}
&4 & & &$T_{\langle-12,0\rangle}$ &$I$ & & & &$T_{\langle84,0\rangle}$ &$T_{\langle96,0\rangle}$ & & & & & & & & & & & & & & & & & & & & & & \\\cline{2-33}
&5 & & & & &$I$ & & & & &$T_{\langle96,0\rangle}$ & & & & & & & & & & & & & & & & & & & & & \\\cline{2-33}
&6 &$T_{\langle-96,0\rangle}$ & & & & &$I$ & & & & & & & & & & & &$S_{-1}T_{\langle460,0\rangle}$ & & & &$T_{\langle408,0\rangle}$ & & & & & & & & & \\\cline{2-33}
&7 & &$T_{\langle-96,0\rangle}$ & & &$S_{-1}T_{\langle216,7\rangle}R_x$ & &$I$ & & &$S_{-1}T_{\langle312,7\rangle}R_x$ & & & & & & & & & & & & & & &$T_{\langle408,0\rangle}$ & & & & & & \\\cline{2-33}
&8 & & &$T_{\langle-96,0\rangle}$ &$T_{\langle-84,0\rangle}$ & & & &$I$ &$T_{\langle12,0\rangle}$ & & & & & & & & & & & & & & & & & & & & & & \\\cline{2-33}
&9 & & &$T_{\langle-108,0\rangle}$ &$T_{\langle-96,0\rangle}$ & & & &$T_{\langle-12,0\rangle}$ &$I$ & & & & & & & & & & & & & & & & & & & & & & \\\cline{2-33}
&10 & & & & &$T_{\langle-96,0\rangle}$ & & & & &$I$ & & & & & & & & & & & & & & & & & & & & & \\\cline{2-33}
&11 & & & & & & & & & & &$I$ & & & & & & & & & & & & & & & & & & & & \\\cline{2-33}
&12 & & & & & & & & & & & &$I$ & & & & & & & &$T_{\langle168,5\rangle}$ & & & & & & & & & & & \\\cline{2-33}
&13 & & &$S_{0}T_{\langle54,7\rangle}$ &$S_{0}T_{\langle66,7\rangle}$ & & & &$S_{0}T_{\langle150,7\rangle}$ &$S_{0}T_{\langle162,7\rangle}$ & & & &$I$ & & & & & & & & & & & & & & & & & & \\\cline{2-33}
&14 & & & & & & & & & & & & & &$I$ & & & & & & & & & & & & & & & & & \\\cline{2-33}
&15 & & &$S_{0}T_{\langle54,0\rangle}$ &$S_{0}T_{\langle66,0\rangle}$ & & & &$S_{0}T_{\langle150,0\rangle}$ &$S_{0}T_{\langle162,0\rangle}$ & & & & & &$I$ & & & & & & & & & & & & & & & & \\\cline{2-33}
&16 & & &$S_{0}T_{\langle54,3\rangle}$ &$S_{0}T_{\langle66,3\rangle}$ & & & &$S_{0}T_{\langle150,3\rangle}$ &$S_{0}T_{\langle162,3\rangle}$ & & & & & & &$I$ & & &$T_{\langle48,0\rangle}$ & & & & & & & & & & &$S_{-1}T_{\langle720,8\rangle}R_x$ & \\\cline{2-33}
&17 & & & & & & & & & & & & & & & & &$I$ & & & & & & & & & & & & & & \\\cline{2-33}
&18 &$S_{-1}T_{\langle364,0\rangle}$ & & & & &$S_{-1}T_{\langle460,0\rangle}$ & & & & & & & & & & & &$I$ & & & &$S_{-1}T_{\langle868,0\rangle}$ & & & & & & & & & \\\cline{2-33}
&19 & & &$S_{0}T_{\langle54,3\rangle}$ &$S_{0}T_{\langle66,3\rangle}$ & & & &$S_{0}T_{\langle150,3\rangle}$ &$S_{0}T_{\langle162,3\rangle}$ & & & & & & &$T_{\langle-48,0\rangle}$ & & &$I$ & & & & & & & & & & &$S_{-1}T_{\langle768,8\rangle}R_x$ & \\\cline{2-33}
&20 & & & & & & & & & & & &$T_{\langle-168,7\rangle}$ & & & & & & & &$I$ & & & & & & & & & & & \\\cline{2-33}
&21 & & & & & & & & & & & & & & & & & & & & &$I$ & & & & & & & & & & \\\cline{2-33}
&22 &$T_{\langle-504,0\rangle}$ & & & & &$T_{\langle-408,0\rangle}$ & & & & & & & & & & & &$S_{-1}T_{\langle868,0\rangle}$ & & & &$I$ & & & & & & & & & \\\cline{2-33}
&23 & & & & & & & & & & & & & & & & & & & & & & &$I$ & & & & & & & & \\\cline{2-33}
&24 & & & & & & & & & & & & & & & & & & & & & & & &$I$ & & & & & & & \\\cline{2-33}
&25 & &$T_{\langle-504,0\rangle}$ & & &$S_{-1}T_{\langle624,7\rangle}R_x$ & &$T_{\langle-408,0\rangle}$ & & &$S_{-1}T_{\langle720,7\rangle}R_x$ & & & & & & & & & & & & & & &$I$ & & & & & & \\\cline{2-33}
&26 & & & & & & & & & & & & & & & & & & & & & & & & & &$I$ &$T_{\langle12,0\rangle}$ & & & & \\\cline{2-33}
&27 & & & & & & & & & & & & & & & & & & & & & & & & & &$T_{\langle-12,0\rangle}$ &$I$ & & & & \\\cline{2-33}
&28 & & & & & & & & & & & & & & & & & & & & & & & & & & & &$I$ & & & \\\cline{2-33}
&A & & & & & & & & & & & & & & & & & & & & & & & & & & & & &$I$ & & \\\cline{2-33}
&B & & &$S_{0}T_{\langle54,5\rangle}R_x$ &$S_{0}T_{\langle66,5\rangle}R_x$ & & & &$S_{0}T_{\langle150,5\rangle}R_x$ &$S_{0}T_{\langle162,5\rangle}R_x$ & & & & & & &$S_{-1}T_{\langle720,8\rangle}R_x$ & & &$S_{-1}T_{\langle768,8\rangle}R_x$ & & & & & & & & & & &$I$ & \\\cline{2-33}
\multirow{-31}{*}{\rotatebox[origin=c]{90}{{\bfseries From pattern}}}&C & & & & & & & & & & & & & & & & & & & & & & & & & & & & & & &$I$ \\\cline{2-33}
\end{tabularx}
}
\end{sidewaystable}

Figure~\ref{RAVEL-HAYDN-GROUND-TRUTH-CM-GT} shows a {\em chroma--midtime\/} (or CM) representation, in which each point, $\langle t,c\rangle$, represents at least one tone in the score, and where $t$ and $c$ are the midtime in tatums and the chroma, respectively, of the tones represented by the point. Table~\ref{RAVEL-HAYDN-GROUND-TRUTH-CM-IPTG} shows the inter-pattern transformation graph for the patterns indicated in this CM representation and the transformation class $F_{\text{2STRmod12}}$.

The CM representation is useful for finding cases where a pattern is reversed and where individual tones in the pattern are displaced to different octaves. For example, occurrence 18 is mapped onto occurrence 22 in the CM representation by the transformation $S_{-1}T_{\langle868,0\rangle}$, but these two patterns are not related by any transformation in either $F_{\mathrm{2STR}}$ or $F_{\mathrm{2STRmod12}}$ in any of the other types of representation considered above. 

\begin{sidewaysfigure}
\resizebox{\linewidth}{!}{\includegraphics{"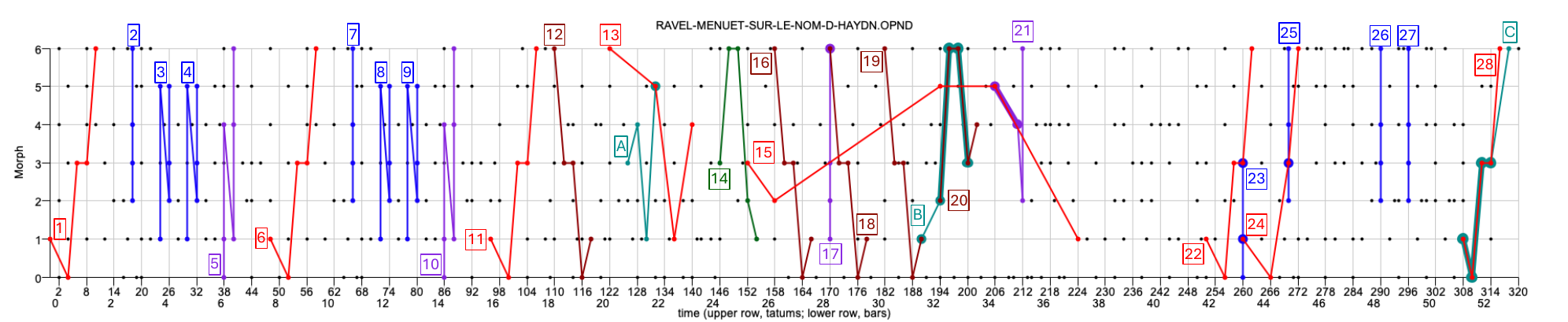"}}
\caption{A {\em morph--onset\/} or MO representation of Ravel's {\em Menuet sur le nom d'Haydn}. Each point represents one or more tones in the score. The $x$-coordinate of a point gives the onset time in tatums (in this case, quavers) of the tones represented by the point, and the $y$--coordinate gives the morph of the tones represented by the point. The occurrences of the HAYDN theme identified in Figs.~\ref{figure-soucy-haydn-occurrences} and \ref{further-occurrences-of-HAYDN-theme} are shown, labelled with the same numbers and letters and coloured using the same colours as in these previous figures. When a point occurs in two patterns, it is drawn twice with different diameters and colours so that its presence in both patterns is visible.}
	\label{RAVEL-HAYDN-GROUND-TRUTH-MO-GT}
\end{sidewaysfigure}

\begin{sidewaystable}
\caption{The inter-pattern transformation graph for the HAYDN occurrences indicated in the MO representation in Fig.~\ref{RAVEL-HAYDN-GROUND-TRUTH-MO-GT} for the transformation class $F_{\mathrm{2STRmod7}}$.}
\label{RAVEL-HAYDN-GROUND-TRUTH-MO-IPTG}
\resizebox{\linewidth}{!}{
\begin{tabularx}{2.15\linewidth}{ll|lllllllllllllllllllllllllllllll}
&&\multicolumn{31}{c}{{\bfseries To pattern}}\\
&&1&2&3&4&5&6&7&8&9&10&11&12&13&14&15&16&17&18&19&20&21&22&23&24&25&26&27&28&A&B&C\\\cline{1-33}
&1 &$I$ &$S_{0}T_{\langle18,3\rangle}$ & & & &$T_{\langle48,0\rangle}$ &$S_{0}T_{\langle66,3\rangle}$ & & & &$T_{\langle96,0\rangle}$ & & & & & &$S_{0}T_{\langle170,5\rangle}R_x$ & & & & &$T_{\langle252,0\rangle}$ & & &$S_{0}T_{\langle270,3\rangle}$ &$S_{0}T_{\langle290,3\rangle}$ &$S_{0}T_{\langle296,3\rangle}$ & & &$T_{\langle190,5\rangle}R_x$ & \\\cline{2-33}
&2 & &$I$ & & & & &$T_{\langle48,0\rangle}$ & & & & & & & & & &$T_{\langle152,2\rangle}R_x$ & & & & & & & &$T_{\langle252,0\rangle}$ &$T_{\langle272,0\rangle}$ &$T_{\langle278,0\rangle}$ & & & & \\\cline{2-33}
&3 & &$S_{0}T_{\langle18,1\rangle}$ &$I$ &$T_{\langle6,0\rangle}$ & & &$S_{0}T_{\langle66,1\rangle}$ &$T_{\langle48,0\rangle}$ &$T_{\langle54,0\rangle}$ & & & & & & & &$S_{0}T_{\langle170,3\rangle}R_x$ & & & & & & & &$S_{0}T_{\langle270,1\rangle}$ &$S_{0}T_{\langle290,1\rangle}$ &$S_{0}T_{\langle296,1\rangle}$ & & & & \\\cline{2-33}
&4 & &$S_{0}T_{\langle18,1\rangle}$ &$T_{\langle-6,0\rangle}$ &$I$ & & &$S_{0}T_{\langle66,1\rangle}$ &$T_{\langle42,0\rangle}$ &$T_{\langle48,0\rangle}$ & & & & & & & &$S_{0}T_{\langle170,3\rangle}R_x$ & & & & & & & &$S_{0}T_{\langle270,1\rangle}$ &$S_{0}T_{\langle290,1\rangle}$ &$S_{0}T_{\langle296,1\rangle}$ & & & & \\\cline{2-33}
&5 & &$S_{0}T_{\langle18,4\rangle}R_x$ & & &$I$ & &$S_{0}T_{\langle66,4\rangle}R_x$ & & &$T_{\langle48,0\rangle}$ & & & & & & &$S_{0}T_{\langle170,2\rangle}$ & & & & & & & &$S_{0}T_{\langle270,4\rangle}R_x$ &$S_{0}T_{\langle290,4\rangle}R_x$ &$S_{0}T_{\langle296,4\rangle}R_x$ & & & & \\\cline{2-33}
&6 &$T_{\langle-48,0\rangle}$ &$S_{0}T_{\langle18,3\rangle}$ & & & &$I$ &$S_{0}T_{\langle66,3\rangle}$ & & & &$T_{\langle48,0\rangle}$ & & & & & &$S_{0}T_{\langle170,5\rangle}R_x$ & & & & &$T_{\langle204,0\rangle}$ & & &$S_{0}T_{\langle270,3\rangle}$ &$S_{0}T_{\langle290,3\rangle}$ &$S_{0}T_{\langle296,3\rangle}$ & & &$T_{\langle142,5\rangle}R_x$ & \\\cline{2-33}
&7 & &$T_{\langle-48,0\rangle}$ & & & & &$I$ & & & & & & & & & &$T_{\langle104,2\rangle}R_x$ & & & & & & & &$T_{\langle204,0\rangle}$ &$T_{\langle224,0\rangle}$ &$T_{\langle230,0\rangle}$ & & & & \\\cline{2-33}
&8 & &$S_{0}T_{\langle18,1\rangle}$ &$T_{\langle-48,0\rangle}$ &$T_{\langle-42,0\rangle}$ & & &$S_{0}T_{\langle66,1\rangle}$ &$I$ &$T_{\langle6,0\rangle}$ & & & & & & & &$S_{0}T_{\langle170,3\rangle}R_x$ & & & & & & & &$S_{0}T_{\langle270,1\rangle}$ &$S_{0}T_{\langle290,1\rangle}$ &$S_{0}T_{\langle296,1\rangle}$ & & & & \\\cline{2-33}
&9 & &$S_{0}T_{\langle18,1\rangle}$ &$T_{\langle-54,0\rangle}$ &$T_{\langle-48,0\rangle}$ & & &$S_{0}T_{\langle66,1\rangle}$ &$T_{\langle-6,0\rangle}$ &$I$ & & & & & & & &$S_{0}T_{\langle170,3\rangle}R_x$ & & & & & & & &$S_{0}T_{\langle270,1\rangle}$ &$S_{0}T_{\langle290,1\rangle}$ &$S_{0}T_{\langle296,1\rangle}$ & & & & \\\cline{2-33}
&10 & &$S_{0}T_{\langle18,4\rangle}R_x$ & & &$T_{\langle-48,0\rangle}$ & &$S_{0}T_{\langle66,4\rangle}R_x$ & & &$I$ & & & & & & &$S_{0}T_{\langle170,2\rangle}$ & & & & & & & &$S_{0}T_{\langle270,4\rangle}R_x$ &$S_{0}T_{\langle290,4\rangle}R_x$ &$S_{0}T_{\langle296,4\rangle}R_x$ & & & & \\\cline{2-33}
&11 &$T_{\langle-96,0\rangle}$ &$S_{0}T_{\langle18,3\rangle}$ & & & &$T_{\langle-48,0\rangle}$ &$S_{0}T_{\langle66,3\rangle}$ & & & &$I$ & & & & & &$S_{0}T_{\langle170,5\rangle}R_x$ & & & & &$T_{\langle156,0\rangle}$ & & &$S_{0}T_{\langle270,3\rangle}$ &$S_{0}T_{\langle290,3\rangle}$ &$S_{0}T_{\langle296,3\rangle}$ & & &$T_{\langle94,5\rangle}R_x$ & \\\cline{2-33}
&12 & &$S_{0}T_{\langle18,3\rangle}$ & & & & &$S_{0}T_{\langle66,3\rangle}$ & & & & &$I$ & &$T_{\langle36,5\rangle}R_x$ & &$T_{\langle48,0\rangle}$ &$S_{0}T_{\langle170,5\rangle}R_x$ &$T_{\langle60,0\rangle}$ &$T_{\langle72,0\rangle}$ &$T_{\langle84,3\rangle}$ & & & & &$S_{0}T_{\langle270,3\rangle}$ &$S_{0}T_{\langle290,3\rangle}$ &$S_{0}T_{\langle296,3\rangle}$ &$S_{-1}T_{\langle426,0\rangle}$ & & & \\\cline{2-33}
&13 & &$S_{0}T_{\langle18,5\rangle}$ & & & & &$S_{0}T_{\langle66,5\rangle}$ & & & & & &$I$ & & & &$S_{0}T_{\langle170,0\rangle}R_x$ & & & & & & & &$S_{0}T_{\langle270,5\rangle}$ &$S_{0}T_{\langle290,5\rangle}$ &$S_{0}T_{\langle296,5\rangle}$ & & & & \\\cline{2-33}
&14 & &$S_{0}T_{\langle18,2\rangle}R_x$ & & & & &$S_{0}T_{\langle66,2\rangle}R_x$ & & & & &$T_{\langle-36,5\rangle}R_x$ & &$I$ & &$T_{\langle12,5\rangle}R_x$ &$S_{0}T_{\langle170,0\rangle}$ &$T_{\langle24,5\rangle}R_x$ &$T_{\langle36,5\rangle}R_x$ &$T_{\langle48,2\rangle}R_x$ & & & & &$S_{0}T_{\langle270,2\rangle}R_x$ &$S_{0}T_{\langle290,2\rangle}R_x$ &$S_{0}T_{\langle296,2\rangle}R_x$ &$S_{-1}T_{\langle462,5\rangle}R_x$ & & & \\\cline{2-33}
&15 & &$S_{0}T_{\langle18,1\rangle}$ & & & & &$S_{0}T_{\langle66,1\rangle}$ & & & & & & & &$I$ & &$S_{0}T_{\langle170,3\rangle}R_x$ & & & & & & & &$S_{0}T_{\langle270,1\rangle}$ &$S_{0}T_{\langle290,1\rangle}$ &$S_{0}T_{\langle296,1\rangle}$ & & & & \\\cline{2-33}
&16 & &$S_{0}T_{\langle18,3\rangle}$ & & & & &$S_{0}T_{\langle66,3\rangle}$ & & & & &$T_{\langle-48,0\rangle}$ & &$T_{\langle-12,5\rangle}R_x$ & &$I$ &$S_{0}T_{\langle170,5\rangle}R_x$ &$T_{\langle12,0\rangle}$ &$T_{\langle24,0\rangle}$ &$T_{\langle36,3\rangle}$ & & & & &$S_{0}T_{\langle270,3\rangle}$ &$S_{0}T_{\langle290,3\rangle}$ &$S_{0}T_{\langle296,3\rangle}$ &$S_{-1}T_{\langle474,0\rangle}$ & & & \\\cline{2-33}
&17 & &$T_{\langle-152,2\rangle}R_x$ & & & & &$T_{\langle-104,2\rangle}R_x$ & & & & & & & & & &$I$ & & & & & & & &$T_{\langle100,2\rangle}R_x$ &$T_{\langle120,2\rangle}R_x$ &$T_{\langle126,2\rangle}R_x$ & & & & \\\cline{2-33}
&18 & &$S_{0}T_{\langle18,3\rangle}$ & & & & &$S_{0}T_{\langle66,3\rangle}$ & & & & &$T_{\langle-60,0\rangle}$ & &$T_{\langle-24,5\rangle}R_x$ & &$T_{\langle-12,0\rangle}$ &$S_{0}T_{\langle170,5\rangle}R_x$ &$I$ &$T_{\langle12,0\rangle}$ &$T_{\langle24,3\rangle}$ & & & & &$S_{0}T_{\langle270,3\rangle}$ &$S_{0}T_{\langle290,3\rangle}$ &$S_{0}T_{\langle296,3\rangle}$ &$S_{-1}T_{\langle486,0\rangle}$ & & & \\\cline{2-33}
&19 & &$S_{0}T_{\langle18,3\rangle}$ & & & & &$S_{0}T_{\langle66,3\rangle}$ & & & & &$T_{\langle-72,0\rangle}$ & &$T_{\langle-36,5\rangle}R_x$ & &$T_{\langle-24,0\rangle}$ &$S_{0}T_{\langle170,5\rangle}R_x$ &$T_{\langle-12,0\rangle}$ &$I$ &$T_{\langle12,3\rangle}$ & & & & &$S_{0}T_{\langle270,3\rangle}$ &$S_{0}T_{\langle290,3\rangle}$ &$S_{0}T_{\langle296,3\rangle}$ &$S_{-1}T_{\langle498,0\rangle}$ & & & \\\cline{2-33}
&20 & &$S_{0}T_{\langle18,0\rangle}$ & & & & &$S_{0}T_{\langle66,0\rangle}$ & & & & &$T_{\langle-84,4\rangle}$ & &$T_{\langle-48,2\rangle}R_x$ & &$T_{\langle-36,4\rangle}$ &$S_{0}T_{\langle170,2\rangle}R_x$ &$T_{\langle-24,4\rangle}$ &$T_{\langle-12,4\rangle}$ &$I$ & & & & &$S_{0}T_{\langle270,0\rangle}$ &$S_{0}T_{\langle290,0\rangle}$ &$S_{0}T_{\langle296,0\rangle}$ &$S_{-1}T_{\langle510,4\rangle}$ & & & \\\cline{2-33}
&21 & &$S_{0}T_{\langle18,6\rangle}R_x$ & & & & &$S_{0}T_{\langle66,6\rangle}R_x$ & & & & & & & & & &$S_{0}T_{\langle170,4\rangle}$ & & & &$I$ & & & &$S_{0}T_{\langle270,6\rangle}R_x$ &$S_{0}T_{\langle290,6\rangle}R_x$ &$S_{0}T_{\langle296,6\rangle}R_x$ & & & & \\\cline{2-33}
&22 &$T_{\langle-252,0\rangle}$ &$S_{0}T_{\langle18,3\rangle}$ & & & &$T_{\langle-204,0\rangle}$ &$S_{0}T_{\langle66,3\rangle}$ & & & &$T_{\langle-156,0\rangle}$ & & & & & &$S_{0}T_{\langle170,5\rangle}R_x$ & & & & &$I$ & & &$S_{0}T_{\langle270,3\rangle}$ &$S_{0}T_{\langle290,3\rangle}$ &$S_{0}T_{\langle296,3\rangle}$ & & &$T_{\langle-62,5\rangle}R_x$ & \\\cline{2-33}
&23 & & & & & & & & & & & & & & & & & & & & & & &$I$ & & & & & & & & \\\cline{2-33}
&24 & &$S_{0}T_{\langle18,3\rangle}$ & & & & &$S_{0}T_{\langle66,3\rangle}$ & & & & & & & & & &$S_{0}T_{\langle170,5\rangle}R_x$ & & & & & & &$I$ &$S_{0}T_{\langle270,3\rangle}$ &$S_{0}T_{\langle290,3\rangle}$ &$S_{0}T_{\langle296,3\rangle}$ & & & & \\\cline{2-33}
&25 & &$T_{\langle-252,0\rangle}$ & & & & &$T_{\langle-204,0\rangle}$ & & & & & & & & & &$T_{\langle-100,2\rangle}R_x$ & & & & & & & &$I$ &$T_{\langle20,0\rangle}$ &$T_{\langle26,0\rangle}$ & & & & \\\cline{2-33}
&26 & &$T_{\langle-272,0\rangle}$ & & & & &$T_{\langle-224,0\rangle}$ & & & & & & & & & &$T_{\langle-120,2\rangle}R_x$ & & & & & & & &$T_{\langle-20,0\rangle}$ &$I$ &$T_{\langle6,0\rangle}$ & & & & \\\cline{2-33}
&27 & &$T_{\langle-278,0\rangle}$ & & & & &$T_{\langle-230,0\rangle}$ & & & & & & & & & &$T_{\langle-126,2\rangle}R_x$ & & & & & & & &$T_{\langle-26,0\rangle}$ &$T_{\langle-6,0\rangle}$ &$I$ & & & & \\\cline{2-33}
&28 & &$S_{0}T_{\langle18,3\rangle}$ & & & & &$S_{0}T_{\langle66,3\rangle}$ & & & & &$S_{-1}T_{\langle426,0\rangle}$ & &$S_{-1}T_{\langle462,5\rangle}R_x$ & &$S_{-1}T_{\langle474,0\rangle}$ &$S_{0}T_{\langle170,5\rangle}R_x$ &$S_{-1}T_{\langle486,0\rangle}$ &$S_{-1}T_{\langle498,0\rangle}$ &$S_{-1}T_{\langle510,3\rangle}$ & & & & &$S_{0}T_{\langle270,3\rangle}$ &$S_{0}T_{\langle290,3\rangle}$ &$S_{0}T_{\langle296,3\rangle}$ &$I$ & & & \\\cline{2-33}
&A & &$S_{0}T_{\langle18,0\rangle}R_x$ & & & & &$S_{0}T_{\langle66,0\rangle}R_x$ & & & & & & & & & &$S_{0}T_{\langle170,5\rangle}$ & & & & & & & &$S_{0}T_{\langle270,0\rangle}R_x$ &$S_{0}T_{\langle290,0\rangle}R_x$ &$S_{0}T_{\langle296,0\rangle}R_x$ & &$I$ & & \\\cline{2-33}
&B &$T_{\langle-190,5\rangle}R_x$ &$S_{0}T_{\langle18,2\rangle}R_x$ & & & &$T_{\langle-142,5\rangle}R_x$ &$S_{0}T_{\langle66,2\rangle}R_x$ & & & &$T_{\langle-94,5\rangle}R_x$ & & & & & &$S_{0}T_{\langle170,0\rangle}$ & & & & &$T_{\langle62,5\rangle}R_x$ & & &$S_{0}T_{\langle270,2\rangle}R_x$ &$S_{0}T_{\langle290,2\rangle}R_x$ &$S_{0}T_{\langle296,2\rangle}R_x$ & & &$I$ & \\\cline{2-33}
\multirow{-31}{*}{\rotatebox[origin=c]{90}{{\bfseries From pattern}}}&C & &$S_{0}T_{\langle18,3\rangle}$ & & & & &$S_{0}T_{\langle66,3\rangle}$ & & & & & & & & & &$S_{0}T_{\langle170,5\rangle}R_x$ & & & & & & & &$S_{0}T_{\langle270,3\rangle}$ &$S_{0}T_{\langle290,3\rangle}$ &$S_{0}T_{\langle296,3\rangle}$ & & & &$I$ \\\cline{2-33}
\end{tabularx}
}
\end{sidewaystable}

Figure~\ref{RAVEL-HAYDN-GROUND-TRUTH-MO-GT} shows a {\em morph--onset-time\/} (or MO) representation, in which each point, $\langle t,m\rangle$, represents at least one tone in the score, and where $t$ and $m$ are the midtime in tatums and the morph, respectively, of the tones represented by the point. Table~\ref{RAVEL-HAYDN-GROUND-TRUTH-MO-IPTG} shows the inter-pattern transformation graph for the patterns indicated in this MO representation and the transformation class $F_{\text{2STRmod7}}$, which is the set of 2-dimensional transformations consisting of an optional scaling parallel to the $x$-axis, followed by a translation in which the $y$ co-ordinate of the translation vector is added $\bmod 7$ to the $y$ co-ordinate of each point, followed by an optional reflection in the $x$-axis (also $\bmod 7$). 

The MO representation is useful for finding cases where a pattern may be augmented or diminished by a non-zero scale factor and then transposed so that it begins on a different key degree and where individual tones in the pattern are displaced to different octaves. For example, in the MO representation, occurrence B is an exact, complete match of occurrence 22 under the transformation $T_{\langle-62,5\rangle}R_x$, but these two patterns are not related by any transformation in either $F_{\mathrm{2STR}}$ or $F_{\mathrm{2STRmod12}}$ in any of the other types of representation considered above.

From Table~\ref{RAVEL-HAYDN-GROUND-TRUTH-MO-IPTG}, it is also evident that this MO representation is useful for discovering cases where a pattern is collapsed into a chord and possibly transposed to a different degree of a scale or inverted. Note that the MO representation allows us to detect that every HAYDN occurrence identified in Figs.~\ref{figure-soucy-haydn-occurrences} and \ref{further-occurrences-of-HAYDN-theme}, apart from the incomplete and possibly erroneous occurrence 23 that lacks a G, can be mapped by transformations in $F_{\text{2STRmod7}}$ onto each of the ``chordal'' occurrences, 2, 7, 17, 25, 26 and 27.

\begin{sidewaysfigure}
\resizebox{\linewidth}{!}{\includegraphics{"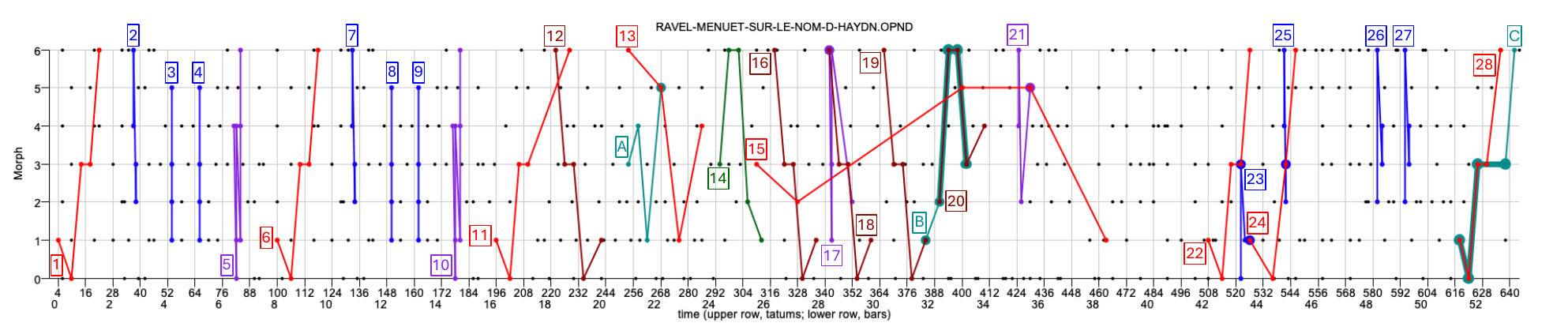"}}
\caption{A {\em morph--midtime\/} or MM representation of Ravel's {\em Menuet sur le nom d'Haydn}. Each point represents one or more tones in the score. The $x$-coordinate of a point gives the midtime in tatums (in this case, semiquavers) of the tones represented by the point, and the $y$--coordinate gives the morph of the tones represented by the point. The occurrences of the HAYDN theme identified in Figs.~\ref{figure-soucy-haydn-occurrences} and \ref{further-occurrences-of-HAYDN-theme} are shown, labelled with the same numbers and letters and coloured using the same colours as in these previous figures. When a point occurs in two patterns, it is drawn twice with different diameters and colours so that its presence in both patterns is visible.}
	\label{RAVEL-HAYDN-GROUND-TRUTH-MM-GT}
\end{sidewaysfigure}

\begin{sidewaystable}
\caption{Inter-pattern transformation graph for occurrences of the HAYDN theme in the MM representation in Fig.~\ref{RAVEL-HAYDN-GROUND-TRUTH-MM-GT} for the transformation class $F_{\mathrm{2STRMod7}}$.}
\label{RAVEL-HAYDN-GROUND-TRUTH-MM-IPTG}
\resizebox{\linewidth}{!}{
\begin{tabularx}{2.25\linewidth}{ll|lllllllllllllllllllllllllllllll}
&&\multicolumn{31}{c}{{\bfseries To pattern}}\\
&&1&2&3&4&5&6&7&8&9&10&11&12&13&14&15&16&17&18&19&20&21&22&23&24&25&26&27&28&A&B&C\\\cline{1-33}
&1 &$I$ & &$S_{0}T_{\langle54,2\rangle}$ &$S_{0}T_{\langle66,2\rangle}$ & &$T_{\langle96,0\rangle}$ & &$S_{0}T_{\langle150,2\rangle}$ &$S_{0}T_{\langle162,2\rangle}$ & & & & &$S_{-1}T_{\langle316,5\rangle}R_x$ & &$S_{-1}T_{\langle340,0\rangle}$ & &$S_{-1}T_{\langle364,0\rangle}$ &$S_{-1}T_{\langle388,0\rangle}$ & & &$T_{\langle504,0\rangle}$ & & & & & & & &$T_{\langle380,5\rangle}R_x$ & \\\cline{2-33}
&2 & &$I$ &$S_{0}T_{\langle54,6\rangle}$ &$S_{0}T_{\langle66,6\rangle}$ &$S_{-1}T_{\langle120,4\rangle}R_x$ & &$T_{\langle96,0\rangle}$ &$S_{0}T_{\langle150,6\rangle}$ &$S_{0}T_{\langle162,6\rangle}$ &$S_{-1}T_{\langle216,4\rangle}R_x$ & & & & & & & & & & & & & & &$T_{\langle504,0\rangle}$ & & & & & & \\\cline{2-33}
&3 & & &$I$ &$T_{\langle12,0\rangle}$ & & & &$T_{\langle96,0\rangle}$ &$T_{\langle108,0\rangle}$ & & & & & & & & & & & & & & & & & & & & & & \\\cline{2-33}
&4 & & &$T_{\langle-12,0\rangle}$ &$I$ & & & &$T_{\langle84,0\rangle}$ &$T_{\langle96,0\rangle}$ & & & & & & & & & & & & & & & & & & & & & & \\\cline{2-33}
&5 & & &$S_{0}T_{\langle54,5\rangle}R_x$ &$S_{0}T_{\langle66,5\rangle}R_x$ &$I$ & & &$S_{0}T_{\langle150,5\rangle}R_x$ &$S_{0}T_{\langle162,5\rangle}R_x$ &$T_{\langle96,0\rangle}$ & & & & & & & & & & & & & & & & & & & & & \\\cline{2-33}
&6 &$T_{\langle-96,0\rangle}$ & &$S_{0}T_{\langle54,2\rangle}$ &$S_{0}T_{\langle66,2\rangle}$ & &$I$ & &$S_{0}T_{\langle150,2\rangle}$ &$S_{0}T_{\langle162,2\rangle}$ & & & & &$S_{-1}T_{\langle412,5\rangle}R_x$ & &$S_{-1}T_{\langle436,0\rangle}$ & &$S_{-1}T_{\langle460,0\rangle}$ &$S_{-1}T_{\langle484,0\rangle}$ & & &$T_{\langle408,0\rangle}$ & & & & & & & &$T_{\langle284,5\rangle}R_x$ & \\\cline{2-33}
&7 & &$T_{\langle-96,0\rangle}$ &$S_{0}T_{\langle54,6\rangle}$ &$S_{0}T_{\langle66,6\rangle}$ &$S_{-1}T_{\langle216,4\rangle}R_x$ & &$I$ &$S_{0}T_{\langle150,6\rangle}$ &$S_{0}T_{\langle162,6\rangle}$ &$S_{-1}T_{\langle312,4\rangle}R_x$ & & & & & & & & & & & & & & &$T_{\langle408,0\rangle}$ & & & & & & \\\cline{2-33}
&8 & & &$T_{\langle-96,0\rangle}$ &$T_{\langle-84,0\rangle}$ & & & &$I$ &$T_{\langle12,0\rangle}$ & & & & & & & & & & & & & & & & & & & & & & \\\cline{2-33}
&9 & & &$T_{\langle-108,0\rangle}$ &$T_{\langle-96,0\rangle}$ & & & &$T_{\langle-12,0\rangle}$ &$I$ & & & & & & & & & & & & & & & & & & & & & & \\\cline{2-33}
&10 & & &$S_{0}T_{\langle54,5\rangle}R_x$ &$S_{0}T_{\langle66,5\rangle}R_x$ &$T_{\langle-96,0\rangle}$ & & &$S_{0}T_{\langle150,5\rangle}R_x$ &$S_{0}T_{\langle162,5\rangle}R_x$ &$I$ & & & & & & & & & & & & & & & & & & & & & \\\cline{2-33}
&11 & & &$S_{0}T_{\langle54,2\rangle}$ &$S_{0}T_{\langle66,2\rangle}$ & & & &$S_{0}T_{\langle150,2\rangle}$ &$S_{0}T_{\langle162,2\rangle}$ & &$I$ & & & & & & & & & & & & & & & & & & & & \\\cline{2-33}
&12 & & &$S_{0}T_{\langle54,2\rangle}$ &$S_{0}T_{\langle66,2\rangle}$ & & & &$S_{0}T_{\langle150,2\rangle}$ &$S_{0}T_{\langle162,2\rangle}$ & & &$I$ & & & & & & & &$T_{\langle168,3\rangle}$ & & & & & & & & & & & \\\cline{2-33}
&13 & & &$S_{0}T_{\langle54,4\rangle}$ &$S_{0}T_{\langle66,4\rangle}$ & & & &$S_{0}T_{\langle150,4\rangle}$ &$S_{0}T_{\langle162,4\rangle}$ & & & &$I$ & & & & & & & & & & & & & & & & & & \\\cline{2-33}
&14 &$S_{-1}T_{\langle316,5\rangle}R_x$ & &$S_{0}T_{\langle54,3\rangle}R_x$ &$S_{0}T_{\langle66,3\rangle}R_x$ & &$S_{-1}T_{\langle412,5\rangle}R_x$ & &$S_{0}T_{\langle150,3\rangle}R_x$ &$S_{0}T_{\langle162,3\rangle}R_x$ & & & & &$I$ & &$T_{\langle24,5\rangle}R_x$ & &$T_{\langle48,5\rangle}R_x$ &$T_{\langle72,5\rangle}R_x$ & & &$S_{-1}T_{\langle820,5\rangle}R_x$ & & & & & & & &$S_{-1}T_{\langle696,0\rangle}$ & \\\cline{2-33}
&15 & & &$S_{0}T_{\langle54,0\rangle}$ &$S_{0}T_{\langle66,0\rangle}$ & & & &$S_{0}T_{\langle150,0\rangle}$ &$S_{0}T_{\langle162,0\rangle}$ & & & & & &$I$ & & & & & & & & & & & & & & & & \\\cline{2-33}
&16 &$S_{-1}T_{\langle340,0\rangle}$ & &$S_{0}T_{\langle54,2\rangle}$ &$S_{0}T_{\langle66,2\rangle}$ & &$S_{-1}T_{\langle436,0\rangle}$ & &$S_{0}T_{\langle150,2\rangle}$ &$S_{0}T_{\langle162,2\rangle}$ & & & & &$T_{\langle-24,5\rangle}R_x$ & &$I$ & &$T_{\langle24,0\rangle}$ &$T_{\langle48,0\rangle}$ & & &$S_{-1}T_{\langle844,0\rangle}$ & & & & & & & &$S_{-1}T_{\langle720,5\rangle}R_x$ & \\\cline{2-33}
&17 & & &$S_{0}T_{\langle54,3\rangle}R_x$ &$S_{0}T_{\langle66,3\rangle}R_x$ &$S_{-0.200000}T_{\langle152.400000,5\rangle}$ & & &$S_{0}T_{\langle150,3\rangle}R_x$ &$S_{0}T_{\langle162,3\rangle}R_x$ &$S_{-0.200000}T_{\langle248.400000,5\rangle}$ & & & & & & &$I$ & & & & & & & & & & & & & & \\\cline{2-33}
&18 &$S_{-1}T_{\langle364,0\rangle}$ & &$S_{0}T_{\langle54,2\rangle}$ &$S_{0}T_{\langle66,2\rangle}$ & &$S_{-1}T_{\langle460,0\rangle}$ & &$S_{0}T_{\langle150,2\rangle}$ &$S_{0}T_{\langle162,2\rangle}$ & & & & &$T_{\langle-48,5\rangle}R_x$ & &$T_{\langle-24,0\rangle}$ & &$I$ &$T_{\langle24,0\rangle}$ & & &$S_{-1}T_{\langle868,0\rangle}$ & & & & & & & &$S_{-1}T_{\langle744,5\rangle}R_x$ & \\\cline{2-33}
&19 &$S_{-1}T_{\langle388,0\rangle}$ & &$S_{0}T_{\langle54,2\rangle}$ &$S_{0}T_{\langle66,2\rangle}$ & &$S_{-1}T_{\langle484,0\rangle}$ & &$S_{0}T_{\langle150,2\rangle}$ &$S_{0}T_{\langle162,2\rangle}$ & & & & &$T_{\langle-72,5\rangle}R_x$ & &$T_{\langle-48,0\rangle}$ & &$T_{\langle-24,0\rangle}$ &$I$ & & &$S_{-1}T_{\langle892,0\rangle}$ & & & & & & & &$S_{-1}T_{\langle768,5\rangle}R_x$ & \\\cline{2-33}
&20 & & &$S_{0}T_{\langle54,6\rangle}$ &$S_{0}T_{\langle66,6\rangle}$ & & & &$S_{0}T_{\langle150,6\rangle}$ &$S_{0}T_{\langle162,6\rangle}$ & & &$T_{\langle-168,4\rangle}$ & & & & & & & &$I$ & & & & & & & & & & & \\\cline{2-33}
&21 & & &$S_{0}T_{\langle54,0\rangle}R_x$ &$S_{0}T_{\langle66,0\rangle}R_x$ & & & &$S_{0}T_{\langle150,0\rangle}R_x$ &$S_{0}T_{\langle162,0\rangle}R_x$ & & & & & & & & & & & &$I$ & & & & & & & & & & \\\cline{2-33}
&22 &$T_{\langle-504,0\rangle}$ & &$S_{0}T_{\langle54,2\rangle}$ &$S_{0}T_{\langle66,2\rangle}$ & &$T_{\langle-408,0\rangle}$ & &$S_{0}T_{\langle150,2\rangle}$ &$S_{0}T_{\langle162,2\rangle}$ & & & & &$S_{-1}T_{\langle820,5\rangle}R_x$ & &$S_{-1}T_{\langle844,0\rangle}$ & &$S_{-1}T_{\langle868,0\rangle}$ &$S_{-1}T_{\langle892,0\rangle}$ & & &$I$ & & & & & & & &$T_{\langle-124,5\rangle}R_x$ & \\\cline{2-33}
&23 & & & & & & & & & & & & & & & & & & & & & & &$I$ & & & & & & & & \\\cline{2-33}
&24 & & &$S_{0}T_{\langle54,2\rangle}$ &$S_{0}T_{\langle66,2\rangle}$ & & & &$S_{0}T_{\langle150,2\rangle}$ &$S_{0}T_{\langle162,2\rangle}$ & & & & & & & & & & & & & & &$I$ & & & & & & & \\\cline{2-33}
&25 & &$T_{\langle-504,0\rangle}$ &$S_{0}T_{\langle54,6\rangle}$ &$S_{0}T_{\langle66,6\rangle}$ &$S_{-1}T_{\langle624,4\rangle}R_x$ & &$T_{\langle-408,0\rangle}$ &$S_{0}T_{\langle150,6\rangle}$ &$S_{0}T_{\langle162,6\rangle}$ &$S_{-1}T_{\langle720,4\rangle}R_x$ & & & & & & & & & & & & & & &$I$ & & & & & & \\\cline{2-33}
&26 & & &$S_{0}T_{\langle54,6\rangle}$ &$S_{0}T_{\langle66,6\rangle}$ & & & &$S_{0}T_{\langle150,6\rangle}$ &$S_{0}T_{\langle162,6\rangle}$ & & & & & & & & & & & & & & & & &$I$ &$T_{\langle12,0\rangle}$ & & & & \\\cline{2-33}
&27 & & &$S_{0}T_{\langle54,6\rangle}$ &$S_{0}T_{\langle66,6\rangle}$ & & & &$S_{0}T_{\langle150,6\rangle}$ &$S_{0}T_{\langle162,6\rangle}$ & & & & & & & & & & & & & & & & &$T_{\langle-12,0\rangle}$ &$I$ & & & & \\\cline{2-33}
&28 & & &$S_{0}T_{\langle54,2\rangle}$ &$S_{0}T_{\langle66,2\rangle}$ & & & &$S_{0}T_{\langle150,2\rangle}$ &$S_{0}T_{\langle162,2\rangle}$ & & & & & & & & & & & & & & & & & & &$I$ & & & \\\cline{2-33}
&A & & &$S_{0}T_{\langle54,1\rangle}R_x$ &$S_{0}T_{\langle66,1\rangle}R_x$ & & & &$S_{0}T_{\langle150,1\rangle}R_x$ &$S_{0}T_{\langle162,1\rangle}R_x$ & & & & & & & & & & & & & & & & & & & &$I$ & & \\\cline{2-33}
&B &$T_{\langle-380,5\rangle}R_x$ & &$S_{0}T_{\langle54,3\rangle}R_x$ &$S_{0}T_{\langle66,3\rangle}R_x$ & &$T_{\langle-284,5\rangle}R_x$ & &$S_{0}T_{\langle150,3\rangle}R_x$ &$S_{0}T_{\langle162,3\rangle}R_x$ & & & & &$S_{-1}T_{\langle696,0\rangle}$ & &$S_{-1}T_{\langle720,5\rangle}R_x$ & &$S_{-1}T_{\langle744,5\rangle}R_x$ &$S_{-1}T_{\langle768,5\rangle}R_x$ & & &$T_{\langle124,5\rangle}R_x$ & & & & & & & &$I$ & \\\cline{2-33}
\multirow{-31}{*}{\rotatebox[origin=c]{90}{{\bfseries From pattern}}}&C & & &$S_{0}T_{\langle54,2\rangle}$ &$S_{0}T_{\langle66,2\rangle}$ & & & &$S_{0}T_{\langle150,2\rangle}$ &$S_{0}T_{\langle162,2\rangle}$ & & & & & & & & & & & & & & & & & & & & & &$I$ \\\cline{2-33}
\end{tabularx}
}
\end{sidewaystable}

Finally, Fig.~\ref{RAVEL-HAYDN-GROUND-TRUTH-MM-GT} shows a {\em morph--midtime\/} (MM) representation of Ravel's {\em Menuet}, in which each point, $\langle t,m\rangle$ represents one or more tones in the score, where $t$ and $m$ represent, respectively, the midtime and the morph of the tones represented by the point. As in the other midtime representations (CPM, MPM and CM), the MM representation allows us to precisely capture the geometric relationship between occurrences related by a retrograde transformation (i.e., a time-scaling by a negative scale factor). For example, in the MM representation, occurrence 22 is mapped completely and exactly onto occurrences 14, 16 and 19 by the transformations $S_{-1}T_{\langle820,5\rangle}R_x$, $S_{-1}T_{\langle844,0\rangle}$ and $S_{-1}T_{\langle892,0\rangle}$, respectively, in the transformation class, $F_{\mathrm{2STRMod7}}$. However, occurrence 22 is not related to occurrences 14, 16 or 19 in any of the other representations considered above by any transformations in $F_{\mathrm{2STR}}$, $F_{\mathrm{2STRMod12}}$ or $F_{\mathrm{2STRMod7}}$. Also, note that, in the MM representation, every occurrence of the HAYDN theme indicated in Figs.~\ref{figure-soucy-haydn-occurrences} and \ref{further-occurrences-of-HAYDN-theme}, apart from the incomplete occurrence 23, can be mapped onto the ``chordal'' occurrences 3, 4, 8 and 9 by some transformation in $F_{\mathrm{2STRMod7}}$.

\section{Summary}

In this paper, we have reviewed the notion of representing music using point sets and argued that such representations are better adapted than sequential representations for matching and discovering patterns in unvoiced, polyphonic music, such as much keyboard music. 
We have also noted that musical transformations such as transposition, inversion, diminution, augmentation and retrograde can be successfully modelled by geometric transformations on sets of points in geometric pitch--time representations that involve combining translation with scaling parallel to and reflection in the time axis. 
We have identified eight specific types of geometric pitch--time representation that use chromatic pitch, morphetic pitch, morph or chroma to represent pitch and either onset time or midtime to represent time. We have illustrated how these different types of representation allow us to precisely characterise different types of musical transformation. For example, by using midtime instead of onset time, we can find and precisely characterise certain retrograde relationships and by using morph and chroma pitch representations we can find and characterise transformations involving octave displacements.
We have also informally introduced the concept of a transformation class and considered three specific examples of such classes, $F_{\mathrm{2STR}}$, $F_{\mathrm{2STRMod7}}$ and $F_{\mathrm{2STRMod12}}$. 
We have introduced the notion of an inter-pattern transformation graph for a set of patterns, $S$, and a transformation class, $F$. Each vertex in such a graph represents a pattern in $S$ and there is an edge in the graph from pattern $P_1$ to pattern $P_2$ if and only if $P_1$ can be mapped exactly and completely onto $P_2$ by one or more transformations in $F$. Each edge in the graph can be labelled with the set of transformations that map the pattern represented by the edge's starting vertex onto the pattern represented by its ending vertex. In this paper, these graphs are represented as tables.
Finally, we have shown, with the aid of such graphs, that the musical relationships between the complete occurrences of the HAYDN theme in Ravel's {\em Menuet sur le nom d'Haydn\/} can be precisely described in terms of transformations in $F_{\mathrm{2STR}}$, $F_{\mathrm{2STRMod7}}$ and $F_{\mathrm{2STRMod12}}$ within one or more of the eight types of pitch--time representation considered.

\appendix
\section{Reproducing the figures and tables in this paper}

In this appendix, we describe how to replicate the pitch--time representations in Figs.~\ref{RAVEL-HAYDN-GROUND-TRUTH-CPO-GT} to \ref{RAVEL-HAYDN-GROUND-TRUTH-MM-GT} and 
the inter-pattern transformation graphs in Tables~\ref{RAVEL-HAYDN-GROUND-TRUTH-CPO-IPTG} to \ref{RAVEL-HAYDN-GROUND-TRUTH-MM-IPTG}.

The inter-pattern transformation graphs can be reproduced using the following procedure:
\begin{enumerate}
\item Ensure that you have a recent Java Runtime or Java SDK installed. You can obtain one here: 
\begin{center}
{\fontsize{8.5pt}{13.5pt}\selectfont \url{https://www.oracle.com/java/technologies/downloads/}}
\end{center}
\item Install a recent Eclipse IDE for Java Developers, available here: 
\begin{center}
{\fontsize{8.5pt}{13.5pt}\selectfont \url{https://www.eclipse.org/downloads/packages/}}
\end{center}
\item Download and unzip the version of the OMNISIA Git repository, here: 
\begin{center}
{\fontsize{8pt}{13.5pt}\selectfont \url{https://github.com/chromamorph/omnisia/releases/tag/GAMABookChapter1-20261001}}
\end{center}
(It is approximately 1 GB.)
\item Launch Eclipse and create a new workspace using the folder containing the downloaded OMNISIA repository. Import all the projects in the downloaded OMNISIA repository into this workspace. Remember to ``Search for nested projects''.
\item Open the file {\ttfamily GAMABookChapter1IPTGs.java} in Eclipse. This file is inside the {\ttfamily com.chromamorph.maxtranpatsjava} package, which is in the {\ttfamily src} folder inside the {\ttfamily MaxTranPatsJava} project folder.
\item Run this file by pressing the green button with a white arrowhead on the toolbar in Eclipse. This will compute the eight inter-pattern transformation graphs and store them as \LaTeX\ files that can be directly input into a \LaTeX\ source file. After the program has run, the files will be found in the folder, 
\begin{center}
 {\ttfamily\fontsize{8.5pt}{13.5pt}\selectfont MaxTranPatsJava$/$output$/$GAMABookChapter1$/$IPTGs$/<$current\_date\_and\_time$>$}
\end{center} 
Ignore any messages that warn you that there are errors in other projects or packages.
\end{enumerate}
The inter-pattern transformation graphs are generated from the file, 
\begin{center}
{\ttfamily\fontsize{8.5pt}{13.5pt}\selectfont data$/$gama$/$chapter1$/$RAVEL-HAYDN-GROUND-TRUTH.gt}
\end{center}
using the parallel, fork/join algorithm described in \cite{MeredithMML2024,MeredithMML2024ProceedingsPaper}.

The pitch--time representations in Figs.~\ref{RAVEL-HAYDN-GROUND-TRUTH-CPO-GT} to \ref{RAVEL-HAYDN-GROUND-TRUTH-MM-GT} can be (partially) reproduced by running eight scripts, named {\ttfamily GAMABookChapter1RavelMenuetXXX.java}, where {\ttfamily XXX} is one of the following: {\ttfamily CPO}, {\ttfamily CPM}, {\ttfamily MPO}, {\ttfamily MPM}, {\ttfamily CO}, {\ttfamily CM}, {\ttfamily MO} or {\ttfamily MM}. These scripts can be found in the same directory as the {\ttfamily GAMABookChapter1IPTGs.java} source file mentioned above. These scripts generate {\ttfamily .png} files that are saved in folders named \begin{center}
{\ttfamily\fontsize{8.5pt}{13.5pt}\selectfont output$/$GAMABookChapter1$/$pitchtimereps$/<$current\_date\_and\_time$>$}
\end{center}
The images generated do not have the pattern numbers indicated in the figures in this paper, as these were added manually by the author. The scripts use the following OPND encoding of Ravel's {\em Menuet}:
\begin{center}
{\ttfamily\fontsize{8.5pt}{13.5pt}\selectfont data$/$gama$/$chapter1$/$RAVEL-MENUET-SUR-LE-NOM-D-HAYDN.OPND}
\end{center}


\end{document}